\documentclass{aa}
\usepackage{graphicx}
\begin{document}

\title{Optical and radio behaviour of the BL Lacertae object
\object{0716+714}\thanks{Tables 3--7 are only available in electronic
form at the CDS via
anonymous ftp to cdsarc.u-strasbg.fr (130.79.125.5)
or via http://cdsweb.u-strasbg.fr/Abstract.html}
}


   \author{C.~M.~Raiteri\inst{1}
          \and M.~Villata\inst{1}
          \and G.~Tosti\inst{2}
          \and R.~Nesci\inst{3}
          \and E.~Massaro\inst{3}
          \and M.~F.~Aller\inst{4}
          \and H.~D.~Aller\inst{4}
          \and H.~Ter\"asranta\inst{5}
          \and O.~M.~Kurtanidze\inst{6,7,8}
          \and M.~G.~Nikolashvili\inst{6}
          \and M.~A.~Ibrahimov\inst{9,10}
          \and I.~E.~Papadakis\inst{11,12}
          \and T.~P.~Krichbaum\inst{13}
          \and A.~Kraus\inst{13}
          \and A.~Witzel\inst{13}
          \and H.~Ungerechts\inst{14}
          \and U.~Lisenfeld\inst{14,}\thanks{\emph{Present address:}
Instituto de Astrof\'{\i}sica de Andaluc\'{\i}a (CSIC),
c/ Camino Bajo de Hu\'etor 24, Apartado 3004, 18080 Granada, Spain}
          \and U.~Bach\inst{13}
          \and G.~Cim\`o\inst{13}
          \and S.~Ciprini\inst{2}
          \and L.~Fuhrmann\inst{13}
          \and G.~N.~Kimeridze\inst{6}
          \and L.~Lanteri\inst{1}
          \and M.~Maesano\inst{15}
          \and F.~Montagni\inst{15}
          \and G.~Nucciarelli\inst{2}
          \and L.~Ostorero\inst{16}
          }

   \institute{Istituto Nazionale di Astrofisica (INAF),
              Osservatorio Astronomico di Torino, Via Osservatorio 20,      
              10025 Pino Torinese (TO), Italy \\        
              \email{raiteri@to.astro.it}
   \and Osservatorio Astronomico, Universit\`a di Perugia, Via B.\ Bonfigli,
        06126 Perugia, Italy \\
        \email{Gino.Tosti@fisica.unipg.it}
   \and Dipartimento di Fisica, Universit\`a di Roma ``La Sapienza'', Piazzale
        Aldo Moro 2, 00185 Roma, Italy \\
        \email{roberto.nesci@uniroma1.it}
   \and Dept.\ of Astronomy, Dennison Bldg., U.\ Michigan, Ann Arbor, MI
   48109, USA \\
         \email{margo@astro.lsa.umich.edu}
   \and Mets\"ahovi Radio Observatory, 02540 Kylm\"al\"a, Finland \\
         \email{hte@alpha.hut.fi}
    \and Abastumani Observatory, 383762 Abastumani, Georgia\\
         \email{okur@kheta.ge}
    \and Astrophysikalisches Institute Potsdam, An der Sternwarte 16, 14482 
     Potsdam, Germany
    \and Landessternwarte Heidelberg-K\"onigstuhl, K\"onigstuhl 12, 69117 
     Heidelberg, Germany
    \and Ulugh Beg Astronomical Institute, Academy of Sciences 
    of Uzbekistan, 33 Astronomical Str., Tashkent 700052, Uzbekistan\\
         \email{mansur@astrin.uzsci.net}
    \and Isaac Newton Institute of Chile, Uzbekistan Branch
    \and Physics Department, University of Crete, 710 03 Heraklion, Crete,
    Greece\\      
         \email{jhep@physics.uoc.gr} 
    \and IESL, Foundation for Research and Technology-Hellas, 711 10
    Heraklion, Crete, Greece
    \and Max-Planck-Institut f\"ur Radioastronomie, Auf dem H\"ugel 69,
              53121 Bonn, Germany\\
         \email{p459kri@mpifr-bonn.mpg.de}    
    \and IRAM, Avd.\ Div.\ Pastora 7NC, 18012 Granada, Spain\\
         \email{ungerechts@iram.es}
    \and Vallinfreda Astronomical Station, Vallinfreda (RM), Italy\\
         \email{f.montagni@libero.it}
    \and Dipartimento di Fisica Generale, Universit\`a di Torino, Via Pietro 
        Giuria 1, 10125 Torino, Italy \\
        \email{ostorero@to.astro.it}
   }

\offprints{C.~M.~Raiteri, \email{raiteri@to.astro.it}}

\date{Received ; accepted}

\abstract{see next page
}

\maketitle
\twocolumn[{\bf Abstract.} Eight optical and four radio  
observatories have been
intensively monitoring the BL Lac object \object{0716+714} in the last years:
4854 data points have been collected in the $UBVRI$ bands since 1994,
while radio light curves extend back to 1978. Many of these data,
which all together constitute the
widest optical and radio database available on this object, 
are presented here for the first time.  
Four major optical
outbursts were observed at the beginning of 1995, in late 1997, at the end of
2000, and in fall 2001. In particular, an exceptional brightening of
2.3 mag in 9 days was detected in the $R$ band just before the BeppoSAX
pointing of October 30, 2000.  
A big radio outburst lasted from early 1998 to the end of 1999.
The long-term trend shown by the optical light curves seems to vary
with a characteristic time scale of about 3.3 years, while a longer period of
5.5--6 years seems to characterize the radio long-term variations.  
In general,
optical colour indices are only weakly correlated with brightness; a
clear spectral steepening trend was observed during at least one long-lasting
dimming phase. Moreover, the optical spectrum
became steeper after $\rm JD \sim 2451000$, the change occurring in
the decaying phase of the late-1997 outburst. 
The radio flux behaviour at different frequencies is similar, 
but the flux variation
amplitude decreases with increasing wavelength. 
The radio spectral index varies with
brightness (harder when brighter), but the radio fluxes seem to be the
sum of two different-spectrum contributions: a steady base level and a
harder-spectrum variable component. Once the base level is removed,
the radio variations appear as essentially achromatic, similarly to
the optical behaviour.
Flux variations at the higher radio frequencies lead the
lower-frequency ones with week--month time scales.
The behaviour of the optical and radio light curves is
quite different, the broad radio outbursts not corresponding in time to the
faster optical ones and the cross-correlation analysis indicating only weak
correlation with long time lags. However, minor radio flux enhancements
simultaneous with the major optical flares can be recognized, which may imply
that the mechanism producing the strong flux increases in the optical band
also marginally affects the radio one. On the contrary, the process
responsible for the big radio outbursts does not seem to affect the optical
emission.

\bigskip]

\section{Introduction}

Understanding blazar variability is one of the major issues of active galactic
nuclei (AGNs) studies. Blazars, i.e.\ BL Lac objects and flat-spectrum radio
quasars, are known to be very active across all the electromagnetic spectrum,
exhibiting flux variations on different time scales, from years down to 
hours or less. It is widely accepted that the variations recognizable
in the blazar
light curves are due to the superposition of a number of components: in
general, one can say that a long-term trend, maybe achromatic
(e.g.\ Ghisellini et al.\ \cite{ghi97}; Villata et al.\ \cite{vil02}), possibly
periodic (e.g.\ Smith \& Nair \cite{smi95}; Raiteri et al.\ \cite{rai01}),
determines the base-level flux oscillations. Faster variations, often
implying spectral changes (e.g.\ Ghisellini et al.\ \cite{ghi97}; Villata et
al.\ \cite{vil00}, \cite{vil02}), are likely caused by a
different physical mechanism.

Verifying the existence of correlations among the flux variations in different
bands is of uttermost importance in order to shed light on the
processes which are at
the origin of the variations themselves. In particular, it is important to
investigate whether a correlation exists between the optical and radio
emissions, which are both ascribed to synchrotron radiation from relativistic
electrons in a plasma jet.

Many studies have been devoted to this main point, but the results for
different sources do not show a common behaviour. On short time scales,
intraday variability (IDV) in radio and optical bands sometimes appeared to be
likely correlated (see Quirrenbach et al.\ \cite{qui91} and  Wagner et al.\
\cite{wag96} for 0716+714; Kraus et al.\ \cite{kra99} for 0235+164). On longer
time scales, when correlations are found, there may be delays among
variations in different bands, changes at the higher frequencies usually
leading those at the lower ones (Tornikoski et al.\ \cite{tor94}; Clements et
al.\ \cite{cle95}; Raiteri et al.\ \cite{rai01}).

In this paper we present the most complete optical and radio light curves of S5
0716+71 available up to now.
Most of the $BVRI$ observations were carried out starting from 1994 by the
blazar monitoring groups of the Perugia and Roma Universities and of the Torino
Observatory, as part of a long-standing collaboration aiming at
studying this BL Lac object. Previous, partial results obtained by the
collaboration have already been published: an analysis of the data  
of the first season can be found in Ghisellini et
al.\ (\cite{ghi97}), while an update of the $R$ light curve up to April 1998
was presented by Massaro et al.\ (\cite{mas99}). Data taken during the
coordinated BeppoSAX and optical observations of November 14, 1996 and
November 7, 1998 were reported by Giommi et al.\ (\cite{gio99}).
Finally, the results of observations performed  during the Whole Earth
Blazar Telescope 
(WEBT; http://www.to.astro/blazars/webt/) campaign of February 1999 appeared in
Villata et al.\ (\cite{vil00}). 
Other optical data have been taken at the Abastumani
Observatory since 1997. The Mount
Maidanak Observatory joined the collaboration during winter 2000--2001, while
optical data simultaneous with the X-ray data taken by the BeppoSAX satellite
on October 30, 2000 were provided by the Skinakas Observatory.

The source 0716+714 is one of the targets of the radio monitoring
  programs at the  Mets\"ahovi Radio Observatory (Ter\"asranta et al.\
\cite{ter98}), at
the University of Michigan Radio Astronomy Observatory (UMRAO; Aller et al.\
\cite{all85}, \cite{all99}), and at the Max-Planck-Institut f\"ur
Radioastronomie in Bonn (Peng et al.\ \cite{pen00}). 
It is also observed 
in the millimetric band at the Instituto de Radioastronomia
Millimetrica in Granada 
(IRAM; Steppe et al.\ \cite{ste88}, \cite{ste92}, \cite{ste93}; 
Reuter et al.\ \cite{reu97}).

A preliminary study of the
cross-correlation between the Mets\"ahovi
plus UMRAO radio data and the optical
ones was presented by Raiteri et al.\ (\cite{rai99}).

The paper is organized as follows: in Sect.\ 2 we review 
previous studies on S5 0716+71 in the optical and radio bands. Optical and
radio observation techniques are described in Sect.\ 3, while light curves are
presented in Sects.\ 4 and 5. Timing analysis and correlations among
bands are discussed in
Sect.\ 6. The conclusions are drawn in Sect.\ 7.


\section{S5 0716+71}

The BL Lacertae object 0716+714 was included in the S5 catalog of the
Strong Source Survey performed at 4.9 GHz (K\"uhr et al.\
\cite{kuh81}). Radio maps reveal a compact core-jet structure and an
extended emission resembling an FR II object (Antonucci et al.\ \cite{ant86};
Gabuzda et al.\ \cite{gab98}). Old estimates of apparent velocity
raised the issue of the existence of superluminal motion in this source
(Gabuzda et al.\ \cite{gab98}); however, a recent determination of proper
motions by Jorstad et al.\ (\cite{jor01}) resulted in values greater than
11--$15 \, h^{-1} \, c$. These latter authors also found that the ejection of VLBA
components may be quasi-periodic, occurring every $\sim 0.7 \, \rm yr$.
The ongoing multifrequency VLBI monitoring (1992--2002; Bach et al.\
\cite{bac02}) by the Bonn group is consistent with the fast-motion scenario
suggested by Jorstad et al.\ (\cite{jor01}). The combination of now more
than 26 VLBI observations at 5--22 GHz results in apparent velocities of
5--$10 \, h^{-1} \, c$, with different components moving at slightly different speeds.
A full analysis of the detailed source kinematics will be given by Bach
et al.\ (\cite{bac03}).

VLA and VLBI observations at 6 cm show
rapid polarization variability, which is probably produced in some compact
feature at about 25 mas from the nucleus (Gabuzda et al.\
\cite{gab00}). Polarization in the optical band was found to be variable on
short time scales too, with possible periods of 12.5, 2.5, and 0.14 days
(Impey et al.\ \cite{imp00}).

Spectroscopic observations of S5 0716+71 have failed to
reveal any feature up to now (Stickel et al.\ \cite{sti93}; Rector \& Stocke
\cite{rec01}), so that the redshift of this source is still undetermined.
However, a number of observational considerations as the starlike appearance
(Stickel et al.\ \cite{sti93}), the absence of a host galaxy, and the small
angular size of the extended halo in radio maps (Wagner et al.\ \cite{wag96})
suggest $z > 0.3$.

Not many photometric data are available in the literature for 0716+714 before
our starting observing date. Biermann et al.\ (\cite{bie81}) reported
variability in magnitude, polarization, and polarization angle. This variability
plus a featureless spectrum shown by the source made them conclude that they
were dealing with a BL Lac object.
The source 0716+714 was included by Beskin et al.\ (\cite{bes85}) in their
photometric study of radio objects with a continuous optical spectrum.
$UBVRI$ photopolarimetric observations were carried out 
by Takalo et al.\ (\cite{tak94}) during two
nights, finding high, variable, and wavelength-dependent polarization.
Dense $BVRI$ monitoring of 0716+714 during a four-week period was performed by
Sagar et al.\ (\cite{sag99}) in 1994. 

Optical IDV was detected by Heidt \& Wagner
(\cite{hei96}), who computed discrete autocorrelation function to search for
possible periodicities in the flux variations and found a period of $4 \, \rm
d$. A study on the optical IDV of 0716+714 during 52 nights was
presented by Nesci et al.\ (\cite{nes02}). They found typical variation rates
of 0.02 mag per hour, and a maximum rising rate of 0.16 mag per hour.

From the analysis of $BVRI$ data taken in 1994--1995, Ghisellini et al.\
(\cite{ghi97}) found no correlation between spectral
index and brightness level in the long-term trend, but were able to detect
a spectral flattening when the flux is higher during rapid flares. From this
they deduced that two processes may be operating in the source: the first one
would cause the achromatic long-term flux variations, while the second one
would be responsible for the fast flux variations implying spectral changes. 
The first process was explained as energy injection
in a large region, which remains stable over at least a few month time scale. 
Two possible interpretations were instead suggested for the fast variations:
a curved trajectory of the relativistic emitting blob or very rapid electron
injection and cooling processes.

Spectral flattening with increasing brightness was also recognized by Villata
et al.\ (\cite{vil00}) in the 72-hour optical light curves obtained
during the WEBT campaign of February
1999. Moreover, by comparison with literature data, a long-term trend
of the spectral index was discovered, implying a steepening of the
optical spectrum and a shift of the synchrotron peak 
(in the $\nu F_\nu$ versus $\nu$ plot) towards the infrared
during the previous five years. The dense sampling achieved during
the WEBT campaign allowed to derive a rate of 0.002 mag per
minute for the steepest flux variations, and an indication of a possible time
delay between variations in the $B$ and $I$ bands, which however must be
shorter than 10 minutes.

An upper limit of $\sim 6$ minutes to the time lag between variations in the
$V$ and $I$ bands was derived by Qian et al.\ (\cite{qia00}) by analyzing
data taken on January 8, 1995. The same authors
also published $BVRI$ light curves from 1994 to 2000 (Qian et al.\
\cite{qia02}), noticing that there may be a roughly 10-day periodicity.

Aller et al.\ (\cite{all85}) presented radio monitoring data (flux density and
linear polarization) at 4.8, 8.0, and 14.5 GHz from the University of Michigan
Radio Astronomy Observatory (UMRAO) in the period 1981--1984. The 1985--1992
data from UMRAO are published in Wagner et al.\ (\cite{wag96}), while radio
fluxes up to 1999 are shown in Raiteri et al.\ (\cite{rai99}), together with 22
and 37 GHz data from the Mets\"ahovi Radio Observatory (see also Ter\"asranta
et al.\ \cite{ter98}) and optical fluxes. 
Radio light curves at 2.7 and 8.1 GHz
were published in Waltman et al.\ (\cite{wal91}). 
The results of radio monitoring at 5, 8.4, and 22 GHz in the
period 1996--1999 with the antennas of Medicina and Noto (Italy) were
presented by Venturi et al.\ (\cite{ven01}).

S5 0716+71 was monitored in the mm band (90--230 GHz) by Steppe et al.\
(\cite{ste88}, \cite{ste92}, \cite{ste93}; see also Reuter et al.\
\cite{reu97}) and Reich et al.\ (\cite{rei93}), showing flux variations of a
factor 2--3 on month time scale and an overall variation of a factor $\sim 6$
in 6 years. Other millimetric (and radio) data can be found in Gear et
al.\ (\cite{gea84}), Edelson (\cite{ede87}), Valtaoja et al.\ (\cite{val92}),
Wiren et al.\ (\cite{wir92}), and Bloom et al.\ (\cite{blo94}). 

The first attempts to monitor the radio IDV of 0716+714 were performed in 1985
at 2.7 GHz using the 100 m radio telescope of the Max-Planck-Institut f\"ur
Radioastronomie, with detection of 5--10\% variations on 1-day time
scale (Heeschen et al.\ \cite{hee87}). 
Similar results were obtained by Quirrenbach et al.\
(\cite{qui89}; see also Quirrenbach et al.\ \cite{qui92}) at 2.7 and 5 GHz, but
without correlation between the two bands. 

In February 1990 a monitoring campaign on simultaneous radio-optical IDV was
organized in order to discriminate whether the fast radio variations can be
ascribed to propagation effects (interstellar scintillation) or if they can
be considered as an intrinsic phenomenon, with all its load of implications. 
The results of the $\sim 4$-week observations at 6 cm and 6500 {\AA}
were presented by Quirrenbach et al.\ (\cite{qui91}): some correlation in the
strong flux variations was found between the two bands, but no strictly
definitive answer. 

Wagner et al.\ (\cite{wag96}) presented an extensive study on rapid variability
of 0716+714 in various energy bands, from radio to X-rays, analyzing
correlation between variations at different frequencies. 
A close correlation was observed through the optical--radio regime, and
possibly between the optical and X-ray bands.


\section{Observations and data reduction}

\subsection{Optical data}

Table 1 gives a list of the optical observatories participating in this study,
together with the telescope size and the number of data collected in different
bands by each of them from 1994 to 2001. A total of 4854 data points was
obtained, which constitute the largest optical dataset on \object{0716+714}
ever published.

\begin{table*}
\caption{Participating optical observatories and number of $UBVRI$ data
collected from 1994 to 2001; the average flux density $<$$F$$>$, the
standard deviation $\sigma$, and the mean fractional variation
$f_{\rm var}$ are also given} 
\begin{tabular}{l c c r r r r r r} 
\hline
Observatory & Label &Diameter
[m]&\multicolumn{1}{c}{$U$}&\multicolumn{1}{c}{$B$}&
\multicolumn{1}{c}{$V$}&\multicolumn{1}{c}{$R$}&\multicolumn{1}{c}{$I$}&
\multicolumn{1}{c}{$N_{\rm tot}$}\\ 
\hline
Greve (Italy)  &GR         & 0.32  &0      & 31  &  33 &  38 &  38 & 140\\ 
Perugia (Italy)  &PG       & 0.40  &0      & 23  & 312 & 522 & 376 & 1233\\ 
Vallinfreda (Italy) &VA    & 0.50  &0      & 214 & 180 & 244 & 194 & 832\\ 
Monte Porzio (Italy) &MP   & 0.70  &0      & 29  &  28 &  37 &  39 & 133\\ 
Abastumani (Georgia) &AB   & 0.70  &0      & 166 & 143 & 490 & 135 & 934\\ 
Torino (Italy)  &TO        & 1.05  &0      & 257 & 163 & 523 &  21 & 964\\ 
Skinakas (Greece)  &SK     & 1.30  &0      & 53  &   0 &  54 &   0 & 107\\
Mt.\ Maidanak (Uzbekistan)&MA & 1.50  &14     & 273 &  14 & 196 &  14 & 511\\ 
\hline
Total                      &  &       &14     &1046 & 873 & 2104& 817 & 4854\\
\hline
$<$$F$$>$ [mJy]                 &  &       &6.35   &8.47 &9.31 &13.18&14.81&\\   
$\sigma$ [mJy]              &  &       &2.06   &3.34 &3.61 &5.63 &5.60 &\\ 
$f_{\rm var}$               &  &       &0.38   &0.39 &0.39 &0.43 &0.38 &\\
\hline
\end{tabular}
\end{table*}

The Perugia frames were taken with the $40 \, \rm cm$ Automatic
Imaging Telescope (AIT) of the Perugia University, mounting a 
$192 \times 165$ pixel CCD cooled by a Peltier stage. 

The University of Roma group has access to three observing facilities at Monte
Porzio, Vallinfreda, and Greve. 
Observations at the $70 \, \rm cm$ f/8.3 TRC70 telescope of Monte Porzio
were performed with a CCD camera
using a back-illuminated SITe $512 \times 512$ chip,
thermoelectrically cooled.
  
The $50 \, \rm cm$ telescope of the Vallinfreda Station is
a Newtonian f/4.5 one, 
equipped with a SBIG ST-6 CCD camera. 
The telescope of Greve is a Newtonian $32 \, \rm cm$ f/4.5 one, 
provided with a CCD camera equal to that mounted on the TRC70.

Data from Torino were taken with the $1.05 \, \rm m$ f/10 REOSC telescope
of the Torino Observatory, equipped with a nitrogen cooled 
$1152 \times 1242$ pixel CCD camera, 
giving an image scale of 0.467 arcsec per pixel.  

The Abastumani Observatory group performs observations in the $BVRI$ bands
with a Meniscus f/3 70 cm telescope, mounting a ST-6 CCD camera.
Details on the Abastumani blazar monitoring program can be found in Kurtanidze
\& Nikolashvili (\cite{kur99}).

Observations at the Mount Maidanak Observatory were done at the
Ritchey-Chr\'etien 1.5 m  f/7.74 telescope equipped with a nitrogen cooled SITe
$2048 \times 800$ pixel CCD, with a  $8.5 \times 3.5$ arcmin field
of view.

$B$ and $R$ frames were taken at the 1.3 m, f/7.7
Ritchey-Chr\'etien telescope of the Skinakas Observatory, with a $1024
\times 1024$ Tektronix CCD camera and a $8 \times 8$ arcmin field of view. 

Data were reduced by either standard packages such as IRAF or reduction
procedures locally developed.

Standard magnitudes in $BVR$ were obtained by using the photometric sequence
published by Villata et al.\ (\cite{vil98a}),  while the $U$ and $I$
data were calibrated
according to Gonz\'alez-P\'erez et al.\ (\cite{gon01}) and Ghisellini et
al.\ (\cite{ghi97}), respectively.

\subsubsection{Data from archival photographic plates}

A few more data in the $B$ band were derived by processing old photographic
plates found in the Torino Observatory plate archive (Table 2). This work
belongs to a more general project aimed at recovering observations
performed in the pre-CCD era. Indeed, a huge number of plates is 
estimated to lie in the
archives of astronomical observatories of all the world, from which important
information on the past flux behaviour of variable sources could be extracted.
 
The Torino plates were processed with a good-quality commercial scanner,
with an optical resolution of $1000 \times 2000$ dpi. Only information
within a $10 \times 10$ arcmin area containing both the source and
the reference stars was considered. Images were treated in a semi-automatic
way, applying the same reduction procedure adopted for the CCD frames. They
were calibrated as standard $B$ frames. This simplified procedure is affected
by errors due to the non-linear plate response to the radiation and to the
fact that the spectral response of the system used does not exactly match a
standard $B$ filter, extending more towards the $U$ band. However, we were able
to obtain satisfactory results (errors of 0.1--0.3 mag) in a very
fast way. 

Data in Table 2 represent moderate/low brightness levels of the source at those
times.

\begin{table}
\caption{$B$ data extracted from Torino Observatory photographic plates}
\begin{tabular}{c c c c}
\hline
Date         &$\rm JD-2440000$ & $B$ & Error\\
\hline
1989 01 27.9   & 7554.4  & 15.55 & 0.13\\
1989 01 30.9   & 7557.4  & 15.66 & 0.19\\
1991 12 29.0   & 8619.5  & 14.48 & 0.15\\
1991 12 29.0   & 8619.5  & 14.63 & 0.15\\
1992 01 02.9   & 8624.4  & 14.93 & 0.24\\
1992 01 02.9   & 8624.4  & 15.03 & 0.33\\
1992 01 06.9   & 8628.4  & 14.77 & 0.11\\
1992 01 06.9   & 8628.4  & 14.87 & 0.22\\
\hline
\end{tabular}
\end{table}

\subsection{Radio band}

Observations at the Mets\"ahovi Radio Observatory are performed with a
13.7 m diameter antenna at 22 and 37 GHz. 
The receivers are dual beam Dicke-type and the flux densities are
calibrated against DR 21. The observing procedure and data reduction are
described in more detail in Ter\"asranta et al.\ (\cite{ter98}).

The UMRAO data are taken with the 26 m paraboloid at 4.8, 8.0, and
14.5 GHz; the observing technique and
reduction procedures are described in Aller et al.\ (\cite{all85}).

Measurements at 1.4, 1.7, 2.7, 5.0, 8.4, 10.7, 15.0, 23, 32, and 43 GHz are
made with the 100 m radio telescope of the Max-Planck-Institut f\"ur
Radioastronomie in Effelsberg. Observational details and calibration
procedures are described, e.g., in Ott et al.\ (\cite{ott94}) and
Peng et al.\ (\cite{pen00}).

Flux densities at 90, 150, and 230 GHz are determined
from observations with the IRAM 30 m telescope at Pico Veleta using procedures,
calibration, and data reduction similar to those described by Steppe
et al.\ (\cite{ste88}) and Reuter et al.\ (\cite{reu97}).

\section{Optical light curves}
\begin{figure*}    
   \includegraphics[width=12cm]{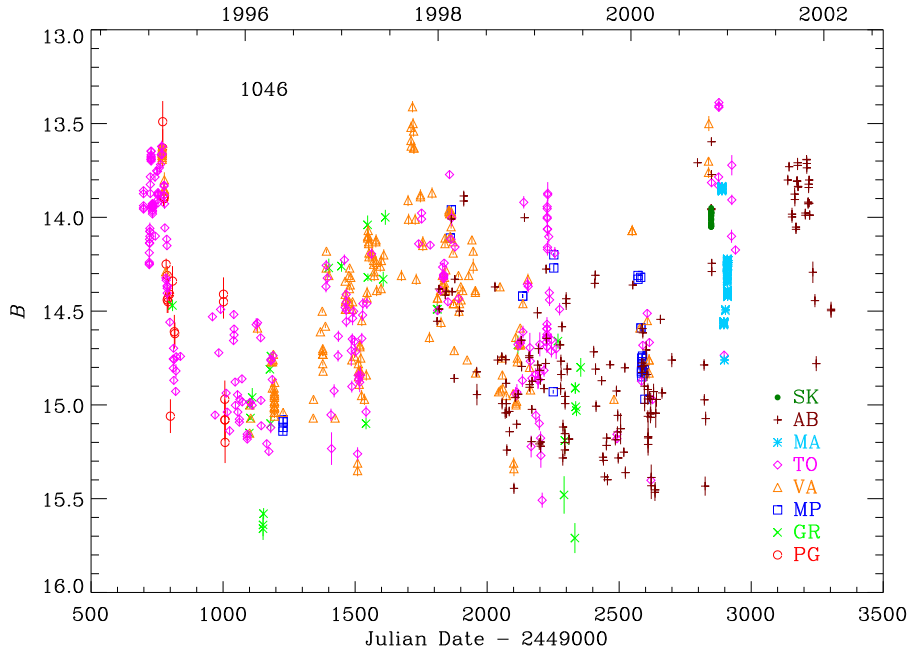}
   \caption{Light curve of S5 0716+71 in the $B$ band}
   \label{btot}
   \end{figure*}

\begin{figure*}
   \includegraphics[width=12cm]{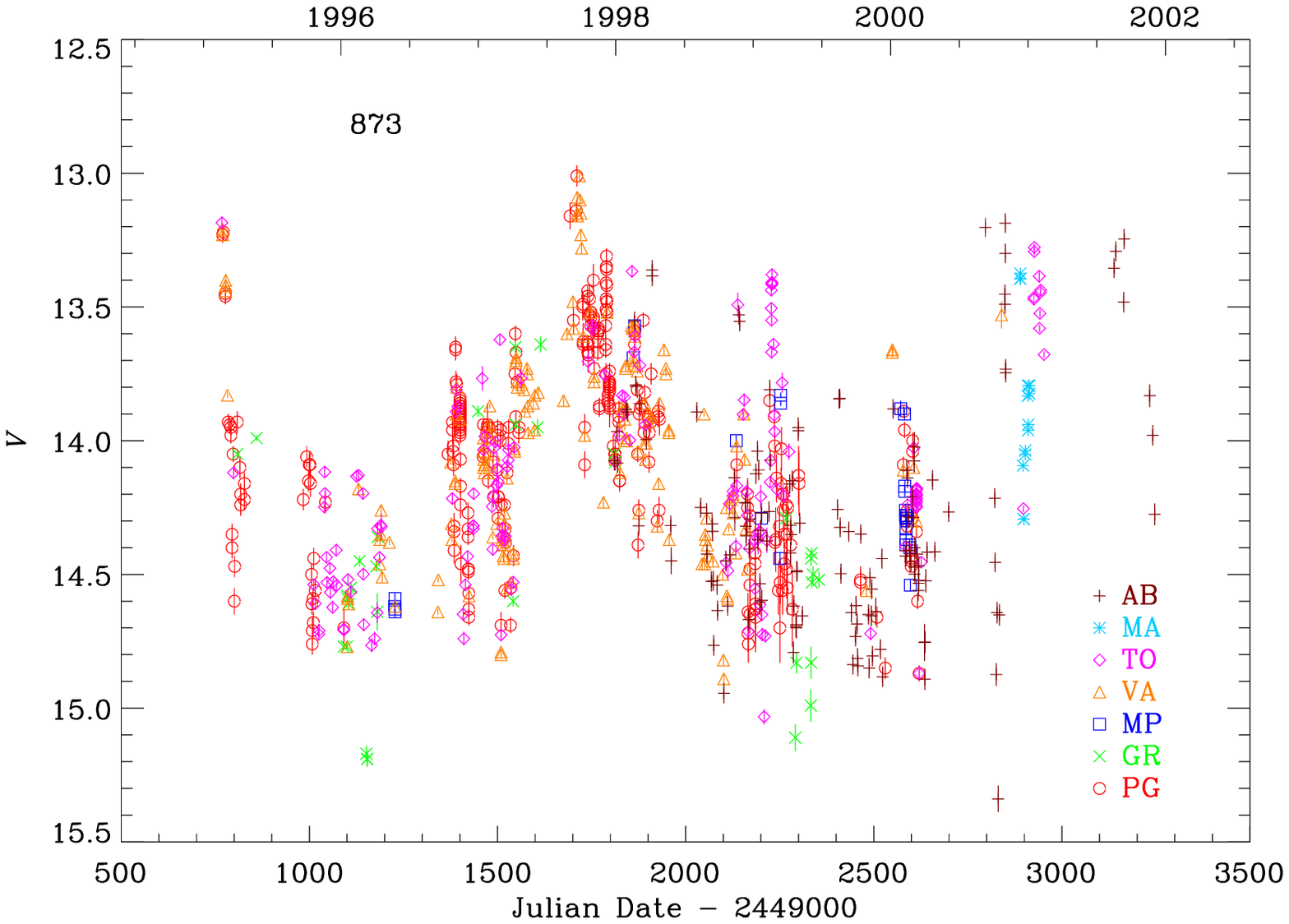}
   \caption{Light curve of S5 0716+71 in the $V$ band}
   \label{vtot}
   \end{figure*}

\begin{figure*}    
   \includegraphics[width=12cm]{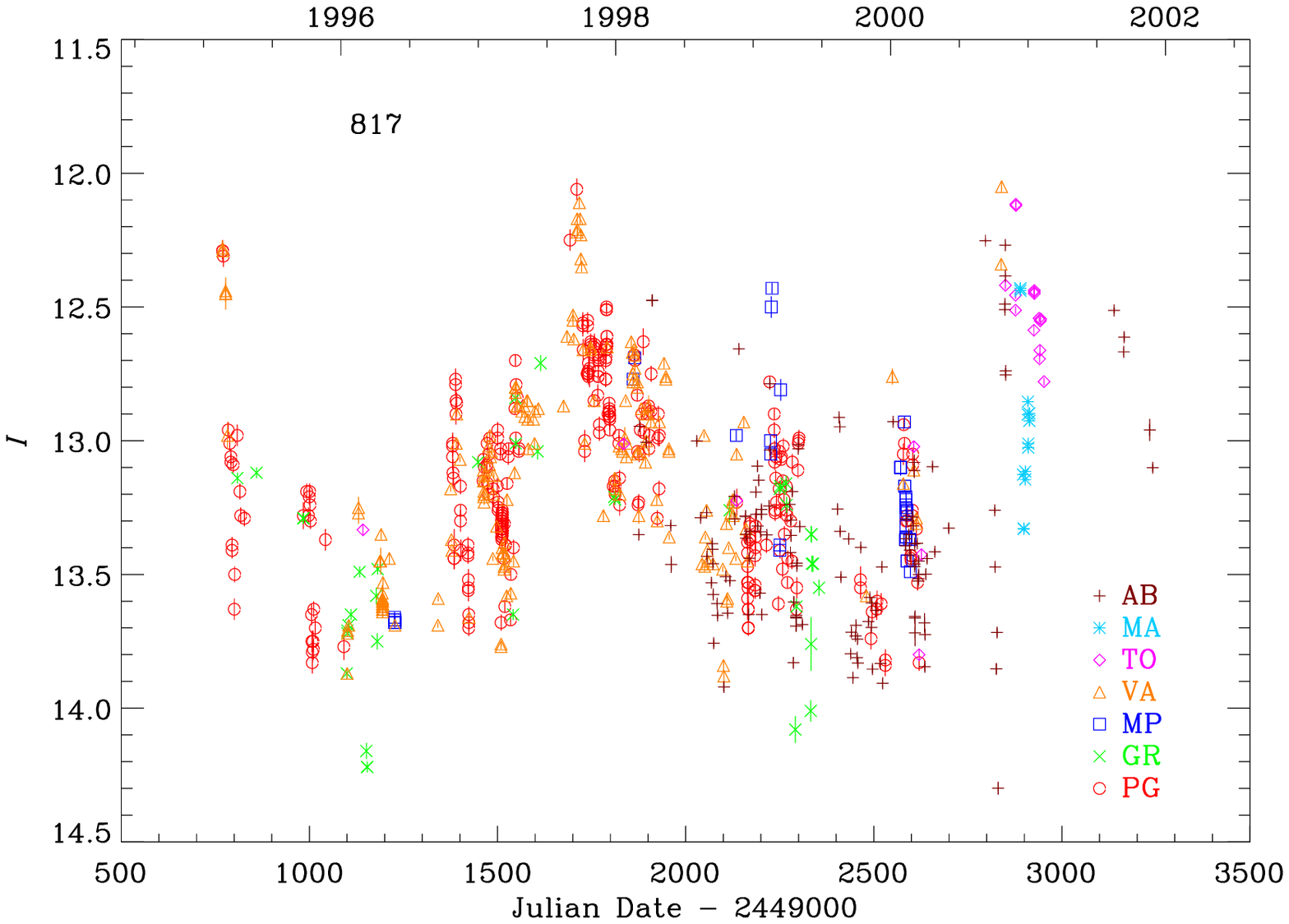}
   \caption{Light curve of S5 0716+71 in the $I$ band}
   \label{itot}
   \end{figure*}

The light curves in the Johnson's $BV$ and Cousins' $RI$ bands are shown in
Figs.\ \ref{btot}--\ref{itot}, where different symbols (and colours) refer to
the various telescopes listed in Table 1. All our $UBVRI$ CCD data are
reported in Tables 3--7, 
available in electronic form.

   \begin{figure*}
   \caption{Light curve of S5 0716+71 in the $R$ band; the BeppoSAX pointing
of October 30, 2000 is indicated by the vertical line; the dashed line
represents a cubic spline interpolation through the binned light curve
(200 days),
excluding the last, undersampled observing season; dotted lines
are obtained by shifting the above spline by $\pm 0.75 \, \rm mag$ in
order to define a $1.5 \, \rm mag$ constant variation area}        
   \label{rtot_spline}
   \end{figure*} 

\begin{table}
\caption{$U$ data}
\begin{tabular}{c c c c}
\hline
$\rm JD-2449000$ & $U$ & Error & Observatory\\
\hline
\end{tabular}
\end{table}

\begin{table}
\caption{$B$ data}
\begin{tabular}{c c c c}
\hline
$\rm JD-2449000$ & $B$ & Error & Observatory\\
\hline
\end{tabular}
\end{table}

\begin{table}
\caption{$V$ data}
\begin{tabular}{c c c c}
\hline
$\rm JD-2449000$ & $V$ & Error & Observatory\\
\hline
\end{tabular}
\end{table}

\begin{table}
\caption{$R$ data}
\begin{tabular}{c c c c}
\hline
$\rm JD-2449000$ & $R$ & Error & Observatory\\
\hline
\end{tabular}
\end{table}

\begin{table}
\caption{$I$ data}
\begin{tabular}{c c c c}
\hline
$\rm JD-2449000$ & $I$ & Error & Observatory\\
\hline
\end{tabular}
\end{table}

The high declination of the source allows a quasi-continuous monitoring during
the year, reducing time gaps due to the solar conjunction. However, technical
reasons as well as bad weather conditions caused poor sampling during the
last observing seasons. 

The lack of dense overlapping among 
the datasets coming from the eight participating observatories
prevents a safe determination of possible magnitude offsets due to the use
of different photometric systems and/or different reduction
procedures. However, 
when data from different telescopes are available during the same night the
agreement is satisfactory. 
  
Light curves appear as the superposition of fast flares lasting a few days
on a modulated base level oscillating on year time scale. Four major
optical outbursts were detected at the beginning of 1995, in late 1997, at
the end of 2000, and in fall 2001. 
In the second-last outburst the source
brightness reached a maximum of $R= 12.54 \pm 0.01$ on $\rm JD=2451877.63$. 

We notice that in the first six observing seasons the long-term
variation amplitude is roughly constant, about 1.5 mag, the trend traced by
the minima being very similar to that traced by the maxima.
This can be seen more easily in  Fig.\
\ref{rtot_spline}, where the dashed line represents the result of a cubic
spline interpolation through the best-sampled $R$-band light curve
(after binning with a 200 day bin size), 
while the dotted lines are obtained by shifting the spline 
by $\pm 0.75 \, \rm mag$. The last, very undersampled observing
season was excluded from the spline interpolation.
By considering the different sampling density in the various observing seasons,
one can say that the constant-amplitude area defined by the dotted curves
contains, in a satisfactory way, all data up to $\rm JD \sim 2451800$; on the
contrary, in the 2000--2001 observing season the variation amplitude is much
larger, exceeding 2 mag. In particular, a
spectacular brightness increase of 2.3 mag in 9 days was detected in October
2000, from $R=14.88 \pm 0.03$ on $\rm JD = 2451830.561$ observed from 
Abastumani,
to $R=12.55 \pm 0.02$ on $\rm JD = 2451839.553$ 
observed from Vallinfreda. This is
one of the strongest variations ever detected not only for this source but
more in general for the blazar class. 
The impressive brightening of the source triggered
the ToO observation of the X-ray satellite BeppoSAX performed on October 30
(see Sect.\ 4.2).

Notice also that the time separation between the $R$ brightness minima
registered on $\rm JD=2450153$ and $\rm JD=2451208$ is about the same 
(1055 days, corresponding to $\sim 2.89$ years) which holds between the maxima
on $\rm JD=2450710$--2450717 and  $\rm JD=2451839$--2451877. 
Actually, a similar time separation ($\sim 2.65$
years) is also found between the peak at $\rm JD=2450710$--2450717 and that at
$\rm JD=2449727$--2449767. 
Although this is obviously a sampling-biased finding, nonetheless it suggests
that a sort of recurrent behaviour may characterize the 0716+714 optical light
curve. Of course, this major point requires a more robust analysis, which will
be performed in Sect.\ 6.

Table 1 also presents some statistics on the optical fluxes: these were
obtained from magnitudes by correcting for a Galactic extinction of
0.081, 0.070, 0.053, 0.045, and 0.032 mag in $UBVRI$, respectively, and
by using the absolute flux densities for zero mag given by Bessell (\cite{bes79}).
Mean fluxes $<$$F$$>$ and standard deviations $\sigma$, in mJy, are reported,
together with the mean fractional variation, defined as
$f_{\rm var}=\sqrt{\sigma^2-\delta^2}/$$<$$F$$>$, where $\delta^2$ is the mean
square uncertainty of the fluxes (Peterson \cite{pet01}). In general,
$f_{\rm var}$ increases with sampling, which makes a comparison among different
bands difficult to perform. One can only say that the mean fractional variation is around 40\%
in all the optical bands.

\subsection{Colour indices}

The existence of spectral changes possibly related to flux variations was
investigated by analyzing colour indices. 
We first derived weighted mean colour indices by
coupling the most precise data (errors not greater than 0.05 mag in $UB$
and 0.04 mag in $VRI$) taken by the same instrument within 30
minutes. The results are shown in Table 8, which also reports the standard
deviation of the sample $\sigma$ and the number of
indices $N$ derived. 
As one can expect, in general the standard deviation increases with the
frequency separation between the two bands, the largest value corresponding to
the $B-I$ index. 

\begin{table}
\caption{Statistics of colour indices} 
\begin{tabular}{l r c r}
\hline
Index  & \multicolumn{1}{c}{Mean} &$\sigma$ &\multicolumn{1}{c}{$N$} \\  
\hline
$U-B$  & $-0.554$ &0.018    &13  \\
$B-V$  & $ 0.434$ &0.062    &444 \\
$B-R$  & $ 0.889$ &0.056    &780 \\ 
$B-I$  & $ 1.382$ &0.097    &356 \\
$V-R$  & $ 0.420$ &0.044    &561 \\
$V-I$  & $ 0.937$ &0.059    &458 \\
$R-I$  & $ 0.524$ &0.052    &522 \\
\hline
\end{tabular}
\end{table}

Then we concentrated on the $B-R$ index, the best sampled one. 
Its behaviour versus time is shown in Fig.\ \ref{colori_jd}
(top panel), where it is compared with the light curve in the $R$ band (bottom
panel). One can see that the colour index is indeed variable, but the
spectral changes do not follow the long-term brightness variations. 

\begin{figure*}
\caption{Temporal evolution of the $B-R$ index (top) compared with the $R$
light curve (bottom); the dashed (red) line in the top panel indicates the
average value $B-R=0.889$}    
\label{colori_jd} 
\end{figure*}

Indeed, when analyzing the data of the
1995--1996 observing season, Ghisellini et al.\ (\cite{ghi97}) found that ``The
colour index correlates with intensity during the rapid flares (the spectrum
is flatter when the flux is higher), but it is rather insensitive to the
long term trends.". The ``achromatic" nature of the long-term flux
oscillations was also recognized by Villata et al.\ (\cite{vil02}) in their
analysis of the optical behaviour of BL Lacertae during the 2000--2001 WEBT
campaign.

If we
calculate the mean $B-R$ for each observing season, we see that it is almost
constant, around 0.84 mag, before $\rm JD \sim 2451000$, while it jumps to
about 0.91 after that date, staying at that level for the subsequent three
seasons (the last season is too undersampled to be included in this analysis).
This transition to a steeper spectral state seems to occur during 
the dimming phase of the big outburst peaking in late 1997. 

\begin{figure}
\resizebox{\hsize}{!}{\includegraphics{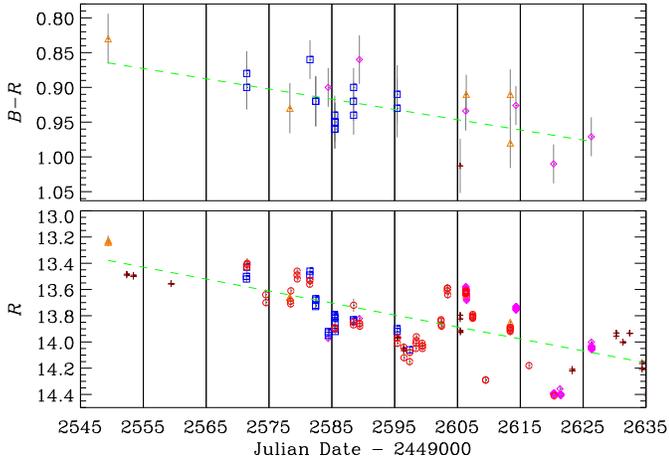}}
\caption{Temporal evolution of the $B-R$ index (top) compared with the $R$
light curve (bottom) in the last part of the sixth observing season; linear
fits and vertical lines have been plotted to guide the
eye; symbols as in Figs.\ \ref{btot}--\ref{itot}}      
\label{colori_jd_obs}  
\end{figure}

On a shorter time scale, a spectral steepening
corresponding to a brightness decrease is also evident in the period
$\rm JD \sim 2451545$--2451630 (see Fig.\ \ref{colori_jd_obs}, where linear
fits have been plotted to guide the eye). 

On even shorter time scales, different behaviours have been found: sometimes a
bluer-when-brighter trend is recognizable, while in some other cases the
opposite is true; there are also cases where magnitude variations do
not imply spectral changes. We think that 
a very dense monitoring with high-precision data is
needed to safely distinguish trends in the short-term spectral behaviour of
this source.

The analysis of the other colour indices leads
to similar results.

\begin{figure}
\resizebox{\hsize}{!}{\includegraphics{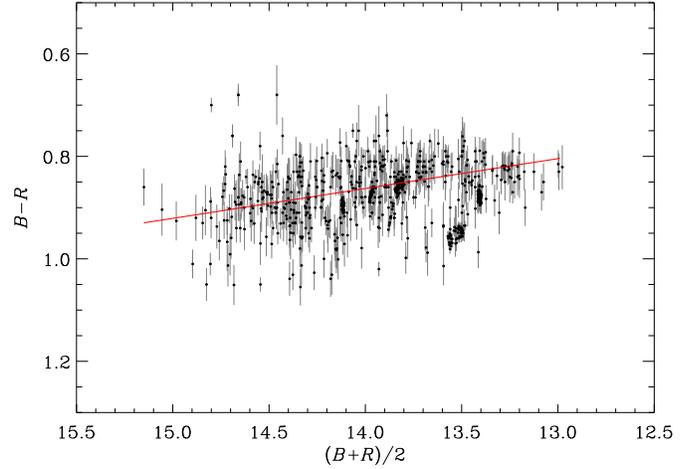}}
\caption{The $B-R$ index as a function of $(B+R)/2$; the linear fit was
obtained after averaging data of the same night }  
\label{bmr_media} 
\end{figure}

\begin{figure}
\resizebox{\hsize}{!}{\includegraphics{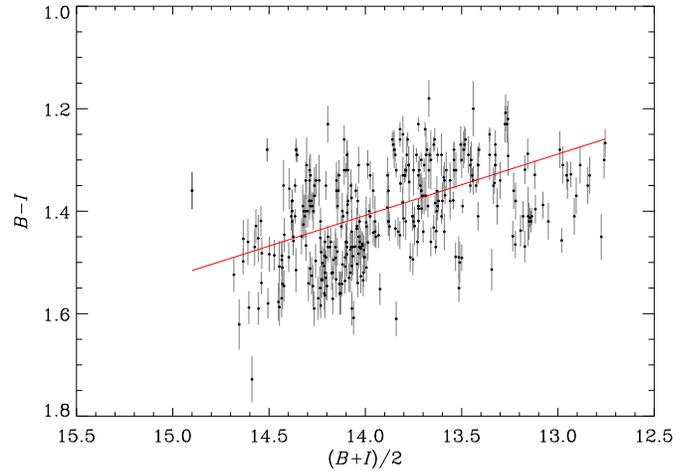}}
\caption{The $B-I$ index as a function of $(B+I)/2$; the linear fit was
obtained after averaging data of the same night } 
\label{bmi_media} 
\end{figure}

The existence of a more general correlation between colour index and
brightness level can be checked by looking at Fig.\ \ref{bmr_media}, where
$B-R$ is plotted versus $(B+R)/2$ in order to minimize the bias introduced by
the dependence of the colour index on the magnitudes that are used to
calculate it (Massaro \& Tr\`evese \cite{mas96}). A linear fit was drawn after
averaging the colour indices obtained during the same night; its slope of
$0.058 \pm 0.003$ and the linear Pearson correlation coefficient
$r=0.15$ suggest a weak correlation between colour index and
brightness. The bluer-when-brighter correlation becomes clearer when plotting
$B-I$ against $(B+I)/2$ (Fig.\ \ref{bmi_media}), because of the larger
frequency separation between $B$ and $I$. In this case the slope of the linear
fit is $0.120 \pm 0.004$ and $r=0.44$.

One can then wonder how much data inhomogeneities due to the fact that we are
working with different datasets may affect our conclusions.
In order to check the influence of systematic offsets we derived
weighted mean $B-R$ values for each observatory (see Table 9, which also gives
$\sigma$, $N$, and the deviation $\Delta$ from the average value 0.889 obtained
by including data from all observatories). As one can see, large deviations
from the average value of 0.889 are found only for the Greve and Perugia mean
values, which are derived from a small number of $BR$ couples. In
particular, all the Perugia indices come from data of the first
observing season, when $B-R$ was intrinsically lower than average (see Fig.\
\ref{colori_jd}). The not negligible deviation of the Skinakas data may be due
to the fact that all of them were taken in the same night, when the
colour index might have been far from its average value. The same sampling
problem also holds for the Mount Maidanak values, all coming from the same
observing season. In this case, however, the deviation from the average value
is quite small.

\begin{table}
\caption{Weighted mean $B-R$ values from each observatory, their
standard deviation $\sigma$, their number $N$, and their deviation from the
weighted mean value of 0.889 reported in Table 8} 
\begin{tabular}{l c c r r}
\hline Observatory    & $<$$B-R$$>$  &$\sigma$ & \multicolumn{1}{c}{$N$}  
& \multicolumn{1}{c}{$\Delta$} \\  
\hline
Greve          & 0.766    & 0.085   & 17 &  $-0.123$\\
Perugia        & 0.779    & 0.037   &  8 &  $-0.110$\\
Vallinfreda    & 0.852    & 0.054   &166 &  $-0.037$\\ 
Monte Porzio   & 0.912    & 0.036   & 23 &  $ 0.023$\\
Abastumani     & 0.896    & 0.069   & 79 &  $ 0.007$\\
Torino         & 0.858    & 0.055   &207 &  $-0.031$\\
Skinakas       & 0.953    & 0.011   & 53 &  $ 0.064$\\
Mt.\ Maidanak & 0.878    & 0.018   &227 &  $-0.011$\\
\hline
\end{tabular}
\end{table}

The deviations $\Delta$ from the average value 0.889 were then used to
``correct"  the colour indices of each observatory, in order to see whether
``normalized" indices may reveal a clearer spectrum-brightness correlation
with respect to that shown by the original data. No shift was assigned to
the Skinakas, Perugia, and Maidanak data, since they refer to very limited time
intervals, as discussed above. The result is very similar to what was obtained
without ``normalization", leading to the conclusion that the possible existence
of systematic colour-index offsets among different observatories does not
affect the general bluer-when-brighter weak correlation.

\subsection{Observations during the BeppoSAX pointing}

The BeppoSAX ToO pointing activated on October 30, 2000 found the source in a
relatively bright optical state. The observations carried out 
at the Skinakas Observatory during that night are shown in Fig.\
\ref{b-r_sk}, where $B$ and $R$ light curves are plotted together with
the $B-R$ colour index. Exposure times of 120 s in $B$ and 60 s in $R$ were
adopted. 

The total magnitude variation was not exceptional, about 0.1 mag. In
general, the
behaviour in the $B$ band is better defined, especially in the first part of
the curve. The $B$ and $R$ light curves follow a similar behaviour, and the
colour index seems to present some correlation with brightness, similar to
that already seen by Villata et al.\ ({\cite{vil00}), i.e.\ with a slight delay
in colour changes (see in particular the central peak of the curves).

\begin{figure}
\resizebox{\hsize}{!}{\includegraphics{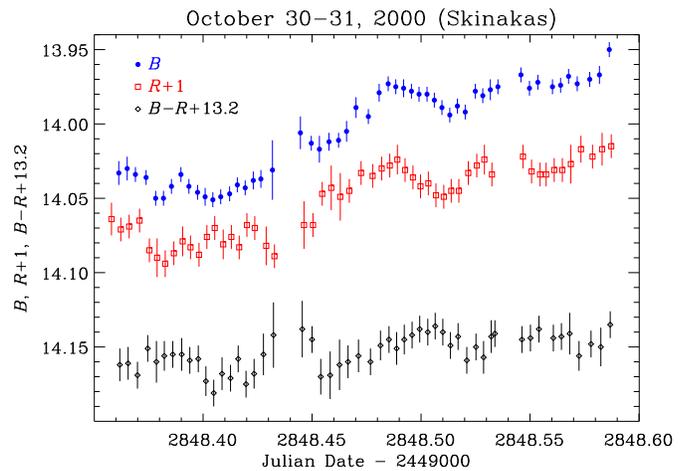}}
\caption{$B$, $R$ ($+1$), and $B-R$ ($+13.2$) curves of October 30--31, 2000,
simultaneous with the BeppoSAX pointing; data are from the Skinakas
Observatory} 
\label{b-r_sk} 
\end{figure}

\subsection{Comparison with the Qian et al.\ (\cite{qia02}) data}

Qian et al.\ (\cite{qia02}) have recently published Johnson's $BV$ and Cousins'
$RI$ data from November 1994 to March 2000 taken with the 1.56
m telescope of the Sheshan Station of the Shangai Astronomical Observatory.
Fig.\ \ref{qian} compares our data in the $R$ and $I$ bands (black dots) with
theirs (grey open circles, green in the electronic version). 

   \begin{figure*}    
   \caption{Comparison between data from this
paper (black dots) and those by Qian et al.\ (\cite{qia02}; grey open circles,
green in the electronic version) in the $R$ (top) and $I$ (bottom) bands}    
\label{qian}    
\end{figure*}

By a first look, 
the Qian et al.\ (\cite{qia02}) data show a total magnitude variation 
much larger than what we detected in our eight observing seasons. 
In particular, one can notice that the most evident disagreement is due to
some data points (their brightness maxima) at a peculiarly stable magnitude
($V \sim 12.5$, $R \sim 12.1$, $I \sim 11.8$), very similar to that of one of
the reference stars used by them. As a consequence, 
they found extremely strong intraday variations, such as that of more than
2 mag in the $R$ band in 21 hours on $\rm JD=2451536$--2451537 and the 1.84 mag
dimming in $I$ band in 40 min a couple of days before (even if there is
disagreement on this latter variation between the printed and the electronic
versions of the data table).
They also report
several other surprising intranight variations, of the order of half a
mag in few minutes.

\section{Radio light curves}

The complete radio flux-density light curves (Jy) obtained by the
  four radio telescopes participating in the present work are plotted in Fig.\
\ref{radio}. They include data taken by the Mets\"ahovi Radio
Observatory (37 and 22 GHz), starting from 1988, observations
  performed by UMRAO (14.5, 8.0, and 4.8 GHz), dating back to 1981, 
data obtained with the Effelsberg radio telescope
(43, 32, 23, 15.0, 10.7, 8.4, 5.0, 2.7, 1.7, and 1.4 GHz)
since 1978, and observations done at the Pico
Veleta radio telescope (230, 150, and 90 GHz), starting from 1983.
Data at 5, 8.4, and 22 GHz  published by Venturi et al.\
(\cite{ven01}) are also shown.

In order to obtain better sampled light curves, 
we combined data at similar frequencies: observations at 22
GHz published by Venturi et al.\ (\cite{ven01}) were added to those at
the same frequency from Mets\"ahovi and to the 23 GHz ones from
Effelsberg; data at 15.0 GHz from Effelsberg were considered together
with those at 14.5 GHz from UMRAO; another combined light curve was
obtained by assembling the 8.0 GHz data from UMRAO with the 8.4 GHz ones
from Effelsberg and Venturi et al.;
finally, 5.0 GHz data from Effelsberg and 5 GHz ones (actually
including 4.9 and 5.0 GHz observations) from Venturi et al.\ were
put together with the 4.8 GHz data from UMRAO.
Moreover, the 230, 150, 90, 43, 37, 32, 2.7, 1.7, and 1.4 GHz data are
shown in Fig.\ \ref{radio} in three panels only, in order to save
space; the 10.7 GHz flux densities are in the same panel of the 8.0--8.4 GHz
light curve.

   \begin{figure*}    
   \centering
   \caption{The complete radio light curves from the Mets\"ahovi Radio
Observatory, UMRAO, Effelsberg and Pico Veleta radio telescopes; in
the 14.5--15.0 GHz plot the (red) dotted line indicates the trend
which was removed in the statistical analysis; 
data from Venturi et al.\ (\cite{ven01}) are also added}      
\label{radio}      	
\end{figure*}

The most extended and best sampled light curves at 15 (14.5--15.0), 8
(8.0--8.4), and 5 (4.8--5.0) GHz seem to suggest an almost
linear decrease of the base-level flux from the starting date until 1995.

The oldest data at 8 (and 5) GHz witness a high state of the source
before 1980, followed by a phase of moderate activity. After 1984 data become
rather sparse up to mid 1993.
In this period outbursts seem to have occurred
in late 1985, 1986--1987, 1988, and 1992.
After that time, an epoch of
essentially low radio brightness can be identified, lasting until early
1998 (see also Fig.\ \ref{radop}). During this period, the radio-flux mean
level is roughly the same at all wavelengths, about 0.5 Jy.  Then the radio
flux started a fast rise, leading to a big outburst, which was characterized
by flaring activity. After more than one year, in mid 1999 the flux was slowly
declining, and at the end of the observing period it was more or less back to
the pre-outburst level.

   \begin{figure*}
   \centering
   \caption{Optical (top panel, mJy) and radio (Jy) flux light curves
   from 1994 to 2001; vertical dashed lines are drawn to guide the eye
   through the major optical flares}        
   \label{radop}
   \end{figure*}

On top of this common behaviour, one has to notice that the variation amplitude
of the radio flux decreases with increasing wavelength, a feature which is
often found in sources of this kind (e.g.\ Aller et al.\ \cite{all85}).
Moreover, there may be a delay of the lower-frequency flux changes with
respect to the higher-frequency ones, which needs to be confirmed by a
cross-correlation analysis (see Sect.\ 6.3).

We notice that there are four maximum points in the 14.5 GHz UMRAO
light curve (red diamonds in Fig. \ref{radio}) that are separated by a time
interval of about 2000--2100 days.
This would suggest a characteristic time
of variability of 5.5--5.7 years for the radio fluxes of 0716+714, which will
be checked in the next section.

Table 10 contains some statistics on the best sampled light curves:
the total number of data
$N$, the mean flux $<$$F$$>$, the standard deviation $\sigma$, and the mean
fractional variation $f_{\rm var}$ already defined for the optical fluxes.
In general, there is a decreasing trend of $<$$F$$>$, $\sigma$,
and $f_{\rm var}$ with decreasing frequency. Notice that, while in the optical
bands $f_{\rm var}$ is almost constant and around 0.4 (see Table 1), in the
radio band it varies substantially with frequency, reaching values much higher
than the optical ones.

This frequency-dependent behaviour seems to be mainly ascribable to
the presence of a flux base level which, unlike the variation
amplitude, does not dim with increasing wavelength.
This suggests that the observed radio flux could be the sum of two
different contributions: the strongly variable flux coming from the
jet would be superposed to a stabler component, possibly coming from
more extended jet regions, or from a steady radio core, or from
elsewhere. 
The two components would have rather different spectra: the variable
component being harder than the base-level one. Thus, the observed
resulting spectrum would vary according to which component
dominates.
This feature is evident in the top panel of 
Fig.\ \ref{alpha}, where
the 5-15 GHz spectral index $\alpha$ ($F_\nu \propto \nu^{- \alpha}$)
is plotted versus the 15 GHz flux. Indeed,
a chromatic behaviour can be recognized,
the spectrum becoming harder when the flux increases, i.e.\ when the
variable component dominates.
In the bottom panel we try to separate the two contributions by
subtracting a base-level flux of 0.25 and 0.35 Jy from
the 15 and 5 GHz data, respectively: an essentially
achromatic behaviour is obtained, points being distributed 
around the mean value $\alpha=-0.25$ (red dotted line). 
One can thus conclude that two components with different spectra and
achromatic behaviour 
may indeed be present in the observed radio fluxes of S5 0716+71,
whose combination gives rise to the observed flux-dependence of the
radio spectral index. According to this interpretation, the long-term  
spectral behaviour of the variable radio component would be consistent 
with the essentially achromatic optical one (see Sect.\ 4.1). 

\begin{figure}
\resizebox{\hsize}{!}{\includegraphics{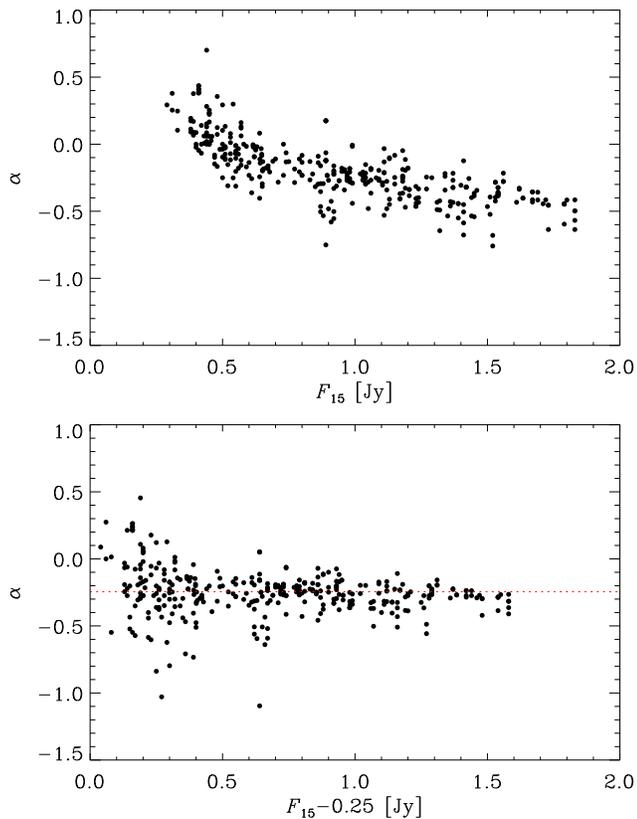}}
\caption{Top: spectral index between the 5 and 15 GHz radio bands
  versus the 15 GHz flux. Bottom: the same as in the top panel after
  subtracting the base levels from the fluxes (see text)}    
\label{alpha} 
\end{figure}

\begin{table}
\caption{Statistics of radio data at different frequencies $\nu$} 
\begin{tabular}{c c c c c}
\hline
$\nu$ [GHz] & $N$ & $<$$F$$>$ [Jy]& $\sigma$ [Jy] & $f_{\rm var}$ \\
\hline
22--23          & 183 & 1.12      & 0.62          & 0.66\\
14.5--15.0      & 527 & 0.88      & 0.42          & 0.61\\
8.0--8.4        & 374 & 0.86      & 0.33          & 0.47\\
4.8--5.0        & 506 & 0.71      & 0.26          & 0.30\\
\hline
\end{tabular}
\end{table}

\section{Statistical analysis}

\subsection{Characteristic time scales of variability}

A number of blazars have revealed a periodic or quasi-periodic
flux behaviour in the optical and/or radio bands, with various
time scales, ranging from a few days to more than 10 years.
Examples of long-term cycles are the $\sim 12$ year period observed in the OJ
287 optical light curve (Sillanp\"a\"a et al.\ \cite{sil88}, \cite{sil96};
Villata et al.\ \cite{vil98b}), the $\sim 7.8$ year period discovered in the
optical light curve of BL Lacertae (Smith \&
Nair \cite{smi95}; Marchenko et al.\ \cite{mar96}), and the $\sim 5.7$ year
period recognized in the radio light curves of AO 0235+16 (Raiteri et al.\
\cite{rai01}). More in general, Smith et al.\ (\cite{smi93}) and Smith \& Nair
(\cite{smi95}) found that many AGNs display smooth, cyclic changes in level on
time scales of years. The BL Lac object 0716+714 was not examined by
them, since it was not included in the Rosemary Hill Observatory AGN
monitoring program they were referring to. As mentioned in Sect.\ 2, hints
for a short-term (a few days) periodic behaviour of the 0716+714 optical light
curve were found by Heidt \& Wagner (\cite{hei96}) and by Qian et al.\
(\cite{qia00}).

In this section we are going to search for characteristic time scales
of variability of S5 0716+71 in both the optical and radio bands.
Beyond the visual inspection of the light curves, a number of statistical
tools will be applied, such as the Discrete Fourier Transform (DFT; see e.g.\
Press et al.\ \cite{pre92}), the Discrete Correlation Function (DCF; Edelson
\& Krolik \cite{ede88}; Hufnagel \& Bregman \cite{huf92}), and the Structure
Function (SF; Simonetti et al.\ \cite{sim85}).
It is common usage to remove a possible linear trend from the datasets
before processing them. 
In the case of 0716+714, no linear trends can be
recognized in the optical light curves, while a smooth flux decrease is
clearly visible in the UMRAO light curves up to $\rm JD \sim 2450000$ (see Fig.
\ref{radio}). 

Application of the DFT leads to a power spectrum, whose  
peaks may indicate the existence of sinusoidal components in the light
curve. We first applied the Fourier analysis to the optical fluxes;
the $R$ flux (mJy) light curve is
displayed  in Fig.\ \ref{radop} (top panel); data taken in 1994 by Sagar et
al.\ (\cite{sag99}) have been added to extend it in time.

The DFT applied to the $R$ fluxes binned over 20 days gives
only one strong signal at a frequency of $8.175 \times 10^{-4} \, \rm
days^{-1}$, corresponding to a period of 1223 days; its false-alarm
probability, which represents the probability that the data values are
independent Gaussian random values, is only $3.08 \times 10^{-7}$ (see Fig.\
\ref{auto_r}, top panel, where the dotted line represents the confidence level
of $10^{-3}$).  

The reliability of this result was then checked by performing the
autocorrelation analysis of the $R$ fluxes by means of the DCF. The original
fluxes were binned over 10 days and the DCF calculated by sampling over
40 days. In this way  spurious effects, which depend on the inverse
square root of the number of points in each DCF bin, are constrained 
within 10\%.
The result is plotted in the middle panel of Fig.\ \ref{auto_r}: the DCF
shows a peak at a time lag of 1160--1240 days. 

Finally, we calculated the SF, which gives an estimate
of the mean flux difference as a function of time separation. 
If a periodic, symmetric signal does exist in the light curve, whose time scale
is $P$, the SF will show alternate maxima and minima at $nP/2$, 
where $n$ are odd and even integers, respectively. The bottom panel of
Fig.\ \ref{auto_r} displays the SF obtained with a bin size of 60 days:
minima are found at 1225 and 2234 days, maxima at 573 and  1770 days, thus
confirming the $\sim 1200$ day characteristic time scale of variability for
the long-term trend of 0716+714 already suggested by the DFT and DCF analyses.

Indeed, a visual inspection of Fig.\ \ref{rtot_spline} (or Fig.\
\ref{radop}) reveals that this time interval is similar to the time
separation between the maximum brightness states observed in 1997 and
late 2000 (see also Sect.\ 4). 

Because of the limited time coverage of
the data used for the present analysis ($\sim 2800$ days) in comparison with
the time scale derived, it goes without saying that further monitoring is
needed to check whether long-term flux changes repeat regularly every $\sim
3.3$ years. 

\begin{figure}
\resizebox{\hsize}{!}{\includegraphics{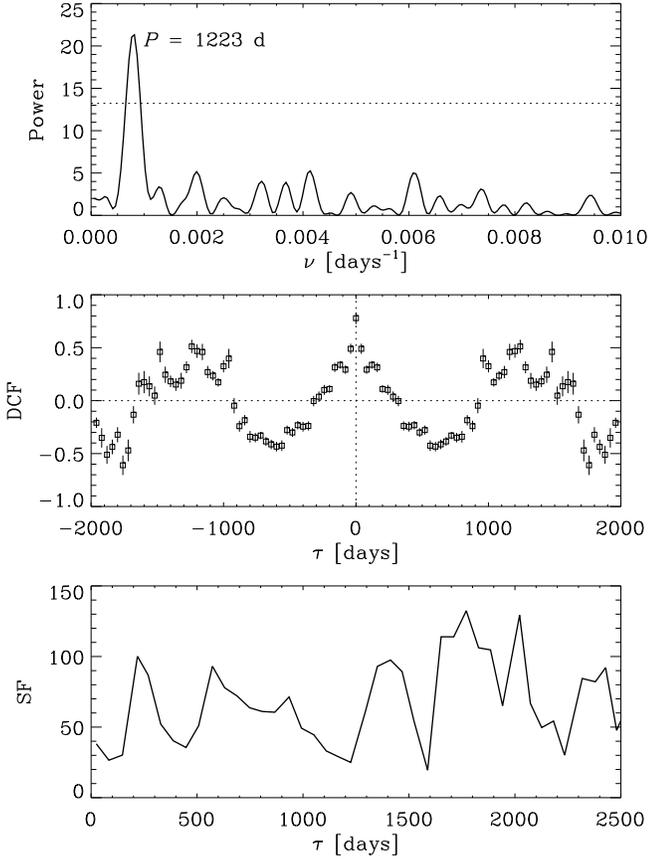}}
\caption{Application of statistical techniques to the $R$ band fluxes of
0716+714: Discrete Fourier Transform periodogram as a function of frequency
(top, the dotted line indicates the level above which the significance
is better than 0.001), Discrete Correlation Function (middle) and Structure
Function (bottom) versus time lag}   
\label{auto_r} 
\end{figure}

We performed the same analysis on the radio light curve obtained
by adding the 15.0 GHz Effelsberg data to the 14.5 GHz UMRAO ones.
This ``15 GHz'' light curve extends for about 7500 days, from early 1981 to
late 2001. 
We applied the DFT, DCF, and SF methods first on the original fluxes (case 1)
and then after subtracting the trend shown in Fig.\ \ref{radio}: a linear
decrease of slope $\sim -1.34 \times 10^{-4} \, \rm Jy \, days^{-1}$ up to
$\rm JD=2450000$, and a constant value of $\sim 0.47$ Jy from then on (case
2). 

The results are shown in Fig.\ \ref{auto_15}: 
black lines in the DFT (top) and SF (bottom) plots refer to case 1
(original data); the (red) dotted lines to case 2 (data corrected for
the trend). In the DCF plot (middle) black squares refer to case 1 and (red)
triangles to case 2.
The DFT analysis reveals that in both cases there are three
  strong signals and two minor ones
  whose significance is better than 0.001 (dotted line in the top
  panel).
In case 1 they correspond to time scales of 6427, 3214, 2045, 1500,
and 703 days; in case 2 the first peak exceeds the time extension of
the light curve and is due to the numerical frequency
oversampling adopted (Press et al.\ \cite{pre92}), 
while the other time scales are 4090, 2368, 738, and 672 days. 
A recurrent time scale of 2045 days would confirm the 2000--2100 day 
periodicity
suggested by the visual inspection of the light curve performed in Sect.\ 5.
The results of the DCF analysis partially confirm the DFT findings,
showing peaks around 2100 days and 4200--4300 days, weak in case 1 and
more pronounced in case 2. The fact that the time lag corresponding to
the latter
peak is twice that of the former one suggests that there is indeed
some recurrent behaviour.
Finally, the SF does not show very pronounced minima for time lags around
2000--2400 days, even if an alternance of maxima and minima with 
$P \sim 2100$--2200 days in case 1 and a couple of deep minima with 
$P \sim 2350$--2700 days in case 2
seem to be consistent with the previous findings.
The large discrepancy between the two cases around $\tau = 6000$ days is due
to the much greater prominence acquired by the 1998--2000 outburst in
case 2. 

\begin{figure}
\resizebox{\hsize}{!}{\includegraphics{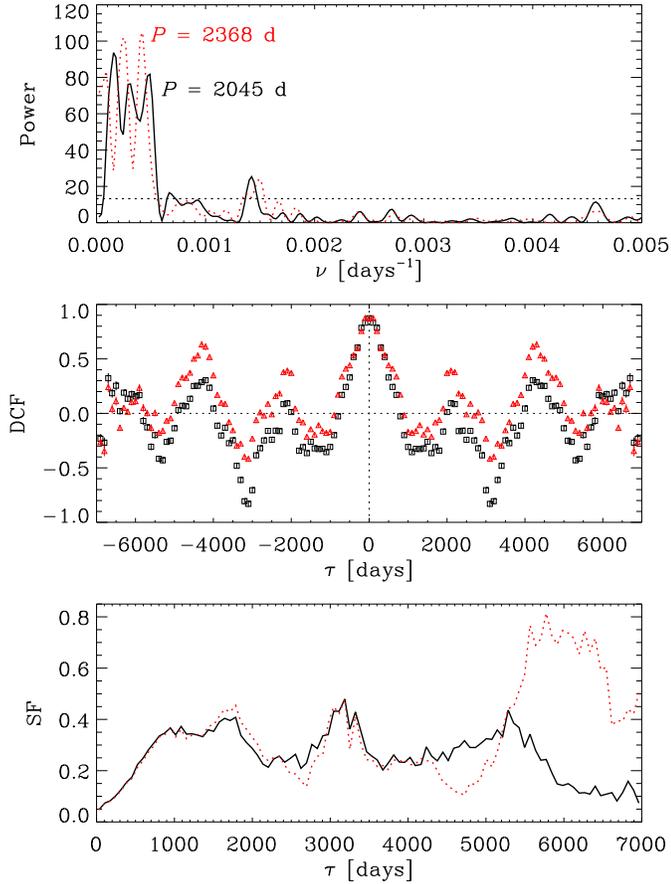}}
\caption{Application of statistical techniques to the 15 GHz light curve
of 0716+714: Discrete Fourier Transform periodogram as a function of frequency
(top, the dotted line indicates the level above which the significance
is better than 0.001), Discrete Correlation Function (middle) and Structure
Function (bottom) versus time lag. Black solid lines and squares refer
to case 1 (original data); (red) dotted lines and triangles to case 2 (data
corrected for the trend shown in Fig.\ \ref{radio})}    
\label{auto_15}  
\end{figure}

The same kind of analysis has been performed on the combined 5 GHz
light curve, containing the 4.8 GHz data from UMRAO, the 5.0 GHz data
from Effelsberg, and the 5 GHz data published by Venturi et al.\
(\cite{ven01}).
The results are shown in Fig.\ \ref{auto_5}. 
No linear trend was subtracted in this case, since the long-term light
curve at this frequency shows only a weak indication of linear flux
decrease (see Fig.\ \ref{radio}).
The periodogram in the top panel of Fig.\ \ref{auto_5} shows a strong
peak
corresponding to a period of 2174 days; peaks at 2100--2200 and
4100--4200 days are evident in the DCF plot (middle panel), while minima
corresponding to similar time scales (and related maxima) 
are found in the SF plot (bottom panel). 

\begin{figure}
\resizebox{\hsize}{!}{\includegraphics{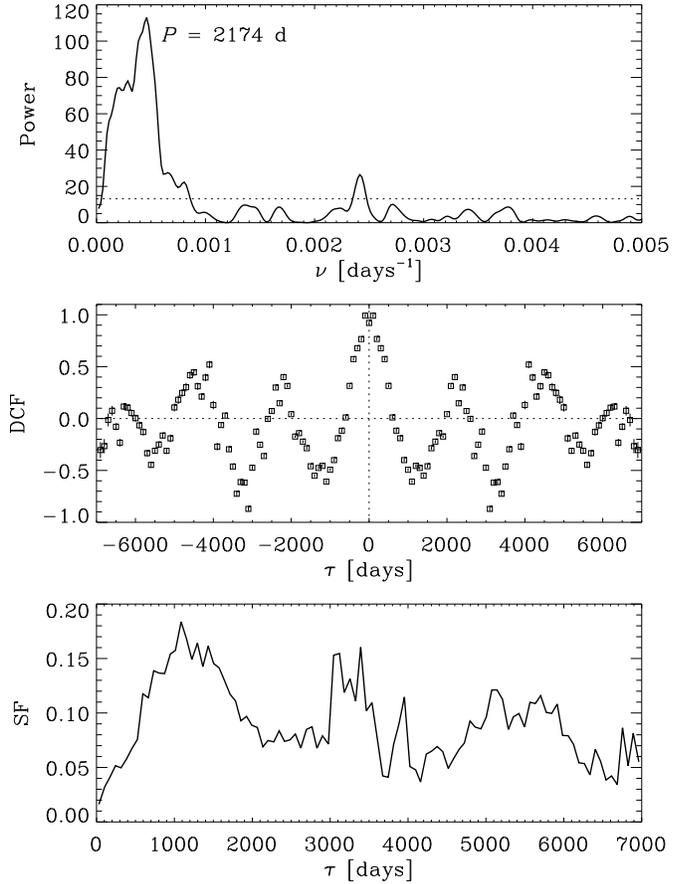}}
\caption{Application of statistical techniques to the 5 GHz light curve
of 0716+714: Discrete Fourier Transform periodogram as a function of frequency
(top, the dotted line indicates the level above which the significance
is better than 0.001), Discrete Correlation Function (middle) and Structure
Function (bottom) versus time lag.}    
\label{auto_5}  
\end{figure}

The same analysis performed on the 8 GHz combined light curve
gives similar, but noisier results.

In conclusion, the analysis of the best-sampled combined datasets
at 15, (8,) and 5 GHz, spanning more than 20 years, 
suggests that a periodic or
quasi-periodic component with time scale of 5.5--6 years
may exist in the radio light curves of 0716+714.
This is the same time scale which characterizes the
recurrence of the major radio outbursts of another BL Lac
object: \object{AO 0235+16} (Raiteri et al.\ \cite{rai01}).

As for the short-term flux variations,
the statistical analysis run on the optical fluxes was not able to reveal any
characteristic time scale, even when fluxes were cleaned from the
contribution of the long-term trend oscillating with a $\sim 3.3$ year time
scale.

\subsection{Optical-radio correlation}

Figure \ref{radop} compares the $R$-band fluxes (mJy, top
panel) with the radio ones (see Fig.\ \ref{radio})
in the period from 1994 to 2001.

An overall inspection of the light curves in Fig.\ \ref{radop}
reveals that the behaviour of the radio fluxes is quite different from the
optical one:
there is not correspondence between the major optical events and the radio
ones. 

As mentioned in the Introduction, the existence of correlation between the
optical and radio long-term trends has been investigated for a number of
blazars: sometimes simultaneous flux variations in the two bands are found; in
other cases correlation appears, but with a time delay, the optical variations
usually leading the radio ones (see e.g.\ Tornikoski et al.\ \cite{tor94};
Clements et al.\ \cite{cle95}; Raiteri et al.\ \cite{rai01}).

We performed cross-correlation analysis between the
optical fluxes and the 15 GHz  combined dataset by means of the DCF; the
result is shown in Fig.\ \ref{dcf_r15} (top panel) and was obtained by first
binning the original data every 10 days, and then choosing a bin size for the
DCF of 80 days. In this way a smooth plot is obtained and at the same time
spurious effects are constrained to less than 5\%. 
A first DCF maximum, indicating a possible
correlation, appears at a time lag $\tau \sim -1000$--$-800$ days.
The peak is found at $\tau=-960$ days and its DCF value is 0.48.
A second, less pronounced maximum is found at $\tau \sim 200$--500
days. The value of the peak is 0.30 and corresponds to a time lag of
400 days.
In the former case, a negative $\tau$ means that radio variations lead
the optical ones, while in the other case the vice versa occurs. The signal at
$\tau \sim -1000$--$-800$ days derives from connecting the radio outbursts
occurred in 1992, early 1995, and 1998--1999 with the optical outbursts
detected in early 1995, late 1997, and end 2000 -- late 2001, respectively,
while that at $\tau \sim 200$--500 days mainly comes from coupling the 
late-1997 optical outburst with the big radio outburst of
1998--1999. In order to obtain a more reliable estimate of the time
lag, it is common usage to calculate the centroid of the DCF,
$\tau_{\rm c}=(\sum_i \tau_i {\rm DCF}_i)/(\sum_i {\rm DCF}_i)$, where sums
run over the points which have a DCF value close to the peak one.
The centroids of the two DCF maxima in Fig.\ \ref{dcf_r15} are $-887$
and 364 days. We notice that, however, the low values of the DCF
corresponding to both peaks indicate weak correlation between optical and
radio variations.

\begin{figure*}
\includegraphics[width=12cm]{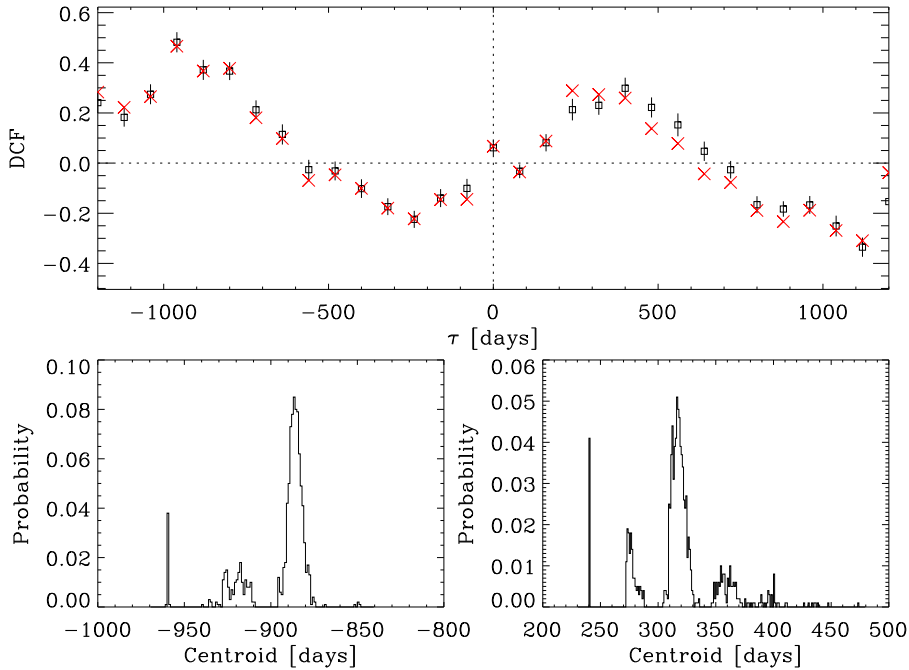}
\caption{Top: DCF between the optical and 15 GHz fluxes (open squares);
red crosses show the mean DCF obtained after averaging 1000 Monte Carlo FR/RSS
realizations. Bottom: CCPD for the negative-$\tau$ (left) and
positive-$\tau$ (right) peaks of the DCF}  
\label{dcf_r15}
\end{figure*}

In order to check uncertainties in cross-correlation lags, we applied the Monte
Carlo technique known as ``flux redistribution/random subset selection"
(FR/RSS; Peterson et al.\ \cite{pet98}). 
Random subsets of the
two datasets to be correlated are selected, redundant points are discarded,
and random Gaussian deviates constrained by the flux errors are added to the
fluxes; the two subsets are then cross-correlated and the resulting DCF
centroid stored. After cross-correlating $N$ Monte Carlo realizations of the
two datasets, an average DCF can be calculated and compared with the original
one. Moreover, a cross-correlation peak (actually, the centroid) distribution
(CCPD) is obtained, from which a measure of the lag uncertainty can be
derived. In this way a test is performed on the influence of both sampling and
errors on the results.

We thus run  FR/RSS processes with $N=1000$ to test the importance of the two
DCF signals in the optical - 15 GHz cross-correlation. 
The average DCF is superposed to the original one in Fig.\ \ref{dcf_r15} (red
crosses): all points but a few ones are inside the original DCF uncertainties.
One interesting feature is that the simulations give more strength to the
lower-$\tau$ part of the optical-leading DCF maximum, evidentiating the
complexity of this signal. The resulting CCPDs are shown in
Fig.\ \ref{dcf_r15} (bottom panels); as for the radio-leading signal,
one can derive that at 76.6\% confidence level the lag is in
between $-896$ and $-877$ days and that at $1 \sigma$ (see Peterson et
al.\ \cite{pet98}) it is $-887^{+7}_{-5}$ days. The situation
regarding the optical-leading signal is more complex: the time lag
estimated from the original DCF falls in a tail of the probability
distribution, while the Monte Carlo simulations indicate a lag 
between 304 and 331 days at 64.6\% confidence level. This means that we
are in the presence of a signal which actually results from the
overlapping of different signals.

In any case, the weakness of the DCF maxima, their complexity, and
especially the very large time delays which they would imply make us
conclude that there is not a reliable radio-optical correlation in
this source.

Another interesting 
feature in Fig.\ \ref{dcf_r15} is that there is only a very weak signal at 
zero time lag, which means that strong events in the optical (radio) band
do not have a strong counterpart in the radio (optical) band. 

However, it is worth performing a deeper visual inspection of the light curves 
in order to see if there is any (although weak) radio
signature simultaneous with the optical outbursts. In order to analyse this
point, dashed vertical lines are drawn in Fig.\ \ref{radop} to guide the eye
through the major optical peaks.
As for the first, double-peaked outburst of
1995, one can only notice that a modest radio peak is
recognizable in the 37--43 GHz light curves, and a slightly delayed
flux enhancement is visible in the 15 and 8 GHz light curves.

The few radio data available around the period when the big optical
outburst of late 1997 occurred witness an enhanced flux with
respect to the surrounding mean level. 

The two less prominent optical peaks detected at the end of 1998 and the
beginning of 1999 occurred during the big radio outburst, so it is
difficult to look for their possible signature in the radio light curves.
However, an enhanced radio flux is visible in correspondence to the 1999 flare
at 22--23 and 15 GHz.

Finally, also the double-peaked optical outburst observed at the end
of 2000 may have a modest radio counterpart, as the millimetric light
curves (second panel from the top in Fig.\ \ref{radop}) seem to suggest, 
even if the radio sampling at that time is rather sparse.

In conclusion, one can say that a modest radio flux increase may correspond to
a major optical outburst. However, we warn that this hypothesis must be
considered with caution because of insufficient data sampling.

By cross-correlating the optical fluxes with the other combined radio light
curves at 5, 8, and 22--23 GHz, one
gets essentially the same results, but the values of the two DCF maxima
increase when cross-correlating higher-frequency radio light curves
since, as seen in Sect.\ 5, the higher the frequency the larger the
variation amplitude. 

Different results are instead obtained when
investigating the optical-millimetric correlation: many signals are
obtained, but the small number of millimetric data prevents a reliable
analysis to be performed. Further intense monitoring at the higher
radio frequencies is a major need in order to better understand the
source multiwavelength behaviour.

\subsection{Radio-radio correlations}

As already noticed, by looking at Figs.\ \ref{radio} and \ref{radop} it is
evident that all the radio bands exhibit a similar behaviour, even if flux
variations become smaller and smaller when going from the highest to the lowest
radio frequencies. This is a rather common feature in blazars; moreover, it
is often found that radio variations at the lower frequencies lag those at the
higher frequencies (see e.g.
Aller et al.\ \cite{all85}, \cite{all99}; Tornikoski et al.\
\cite{tor94}; Raiteri et al.\ \cite{rai01}). In order to investigate this
point, in this section we perform cross-correlation analysis between fluxes at
different radio wavelengths by means of the DCF method. 

The result of cross-correlating the 22--23 GHz  and 15 GHz datasets is
presented in Fig. \ref{dcf_2215} (left panel).  

   \begin{figure}
   \resizebox{\hsize}{!}{\includegraphics{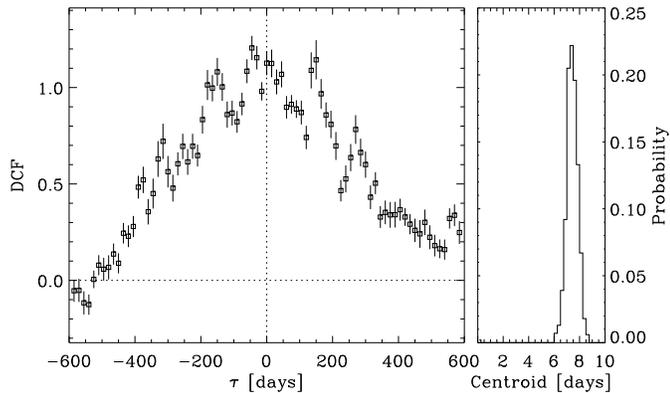}}
   \caption{Left: cross-correlation between the 22--23 and 15 GHz data.
Right: CCPD relative to the central
peak obtained by running 1000 FR/RSS Monte Carlo realizations}         
\label{dcf_2215}    
\end{figure}

The four most
significant maxima are found at time lags of $\sim -150$,
$\sim -50$--$-30$, $\sim 0$--20, and $\sim
150$ days. In particular, the peak at $\tau \sim 0$--20 days suggests that
flux variations at 15 GHz may be delayed with respect to those at
22--23 GHz by some days. 
The lack of a better sampling does not allow us to improve the $\tau$
resolution of the plot without increasing too much spurious effects, which
are now constrained to a few percents. 
Calculation of the centroid for the central peak gives
$\tau_{\rm c}=7$ days. Its CCPD is shown in the right panel
of Fig.\ \ref{dcf_2215}: from it one can derive that at the 99.8\% confidence
level the time delay is 6--9 days, which is significantly different
from zero, i.e.\ we can conclude that flux variations at 15 GHz do lag
those at 22--23 GHz.

An indication of a larger delay is found by cross-correlating the 22--23
GHz data with the 8 GHz ones.
The result is shown in Fig.\ \ref{dcf_228} (left panel): the peak is
found at $\tau=15$ days, but the centroid gives a time lag of 22 days.
The CCPD (right panel) allows one to derive that at $1 \sigma$
$\tau=22 \pm 2$ days.

   \begin{figure}
   \resizebox{\hsize}{!}{\includegraphics{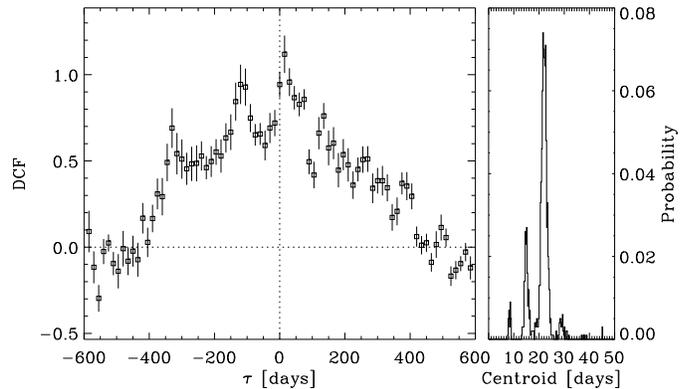}}
   \caption{Left: cross-correlation between the 22--23 and 8
GHz data. Right: CCPD relative to the central
peak obtained by running 1000 FR/RSS Monte Carlo realizations}        
\label{dcf_228}    
\end{figure}

We finally tested the correlation between the 15 and 5 GHz fluxes. 
As can be seen in Fig. \ref{dcf_155} (left
panel) the result is again a delay, the DCF peaking at $\tau=45$
days, while $\tau_{\rm c} = 53$ days. The CCPD confirms the
reliability of 
this delay (see Fig.\ \ref{dcf_155}, right panel): 
the time lag is $\tau=53 \pm 2$ days at 93\% confidence level. 

   \begin{figure}
   \resizebox{\hsize}{!}{\includegraphics{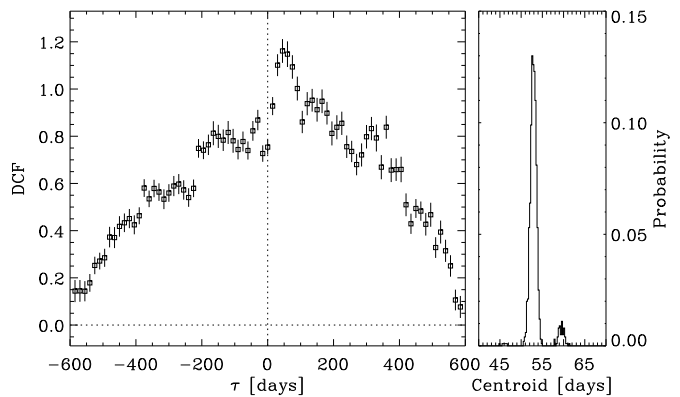}}
   \caption{Left: cross-correlation between the 15 and 5 GHz data. 
Right: CCPD relative to
the central peak obtained by running 1000 FR/RSS Monte Carlo realizations}    
    \label{dcf_155}    
\end{figure}

The above results suggest that the flux variations at
the lower radio frequencies are delayed with respect to those at the higher
frequencies. }


\section{Conclusions}

In this paper we have presented the most complete optical and radio 
light curves of
the BL Lacertae object 0716+714 ever published. 4854 $UBVRI$ data
were taken from eight observatories during eight observing seasons, from 1994
to 2001.
Radio light curves from 1.4 to 230 GHz, most of which
spanning more than 20 years, were obtained by four
radio observatories.

A long-term trend is visible in the optical light curves, on which variations
on shorter time scales are superposed; statistical analysis by means of DFT,
DCF, and SF suggests that the long-term trend may oscillate on a $\sim
3.3$ year
time scale, while no periodicity is found for the fast flux changes. The same
analysis performed on the best-sampled combined datasets at 5, 8,
and 15 GHz indicates that radio fluxes may
have a variability time scale of 5.5--6 years. It is worth noticing
that a similar period was recognized in the radio light curve of another BL Lac
object: AO 0235+16 (Raiteri et al.\ \cite{rai01}).

During the first six observing seasons the variation amplitude of the
optical light curves is roughly constant, about 1.5 mag, but in the
2000--2001 season a dramatic brightening of 2.3 mag in 9 days was
observed, leading to an increase of the variability amplitude.

A constant variability amplitude in magnitudes implies a flux
variation amplitude
which is proportional to the flux level,
 and this can be easily explained in terms
of a variation of the Doppler beaming factor $\delta=[\Gamma(1-\beta \cos
\theta)]^{-1}$, where $\Gamma$ is the Lorentz
factor of the bulk motion of the emitting plasma in the jet
and $\theta$ is the viewing angle.
In this framework, assuming that the intrinsic flux
is relativistically enhanced by a factor $\delta^3$, the maximum oscillation of
the long-term trend identified in Fig.\ \ref{rtot_spline} would imply a
variation of $\delta$ of a factor $\sim 1.3$. 
This change could be due to
either energetic or geometrical reasons, or to a combination of both.
In a paper on the optical variability of BL Lacertae during the May 2000 --
January 2001 WEBT campaign, Villata et al.\ (\cite{vil02}) favoured a
geometrical interpretation. The case of 0716+714 is very similar, since a
variation of a factor 1.3 of $\delta$ can be explained 
by a variation of 
few degrees of the viewing angle, while it would require a noticeable change of
the Lorentz factor.

Colour analysis on the optical light curves reveals only a weak general
correlation between the colour index and the source brightness. In at
least one case a clear spectral steepening was observed during a long-lasting
dimming phase; moreover, the optical spectrum after
$\rm JD \sim 2451000$ is on the average redder than before, 
the change occurring in
the decaying phase of the late-1997 outburst.
On shorter time scales, different spectral behaviours
were found.

Radio light curves at different frequencies have a similar behaviour, but the
flux variation amplitude becomes smaller with decreasing frequency, 
a feature which is often observed in blazars 
(e.g.\ Aller et al.\ \cite{all85}). 
  Unlike the essentially achromatic optical long-term behaviour,
  the radio spectrum becomes harder when the source brightens. However,
  this behaviour can be interpreted as due to the combination of two
  contributions with different spectra: a steady base level and a
  harder-spectrum variable component.
  Both contributions are achromatic,
  but the resulting spectrum varies according to which component dominates. 

Cross-correlation between radio light curves indicates that, 
with a high degree of correlation, flux variations
at lower frequencies lag the higher-frequency ones with time delays
from a few days to several weeks, depending on the frequency separation.
Again, this has already been observed for other blazars and
has been interpreted in various ways. In the framework of a
homogeneous model, where the radiation is produced in a single
homogeneous blob relativistically moving at a small viewing angle, 
lags are interpreted in terms of electron cooling time scales.
On the other hand, in
an inhomogeneous jet the higher synchrotron frequencies are
emitted from the inner, denser parts of the jet, while the lower ones
are emitted from progressively more external regions. In this case, the time
lag is a measure of the distance between emitting regions in the
jet. Indeed, a perturbance travelling down the jet would trigger the
emission of different frequencies at different times. Also in a
rotating helical jet model the jet inhomogeneity causes
time lags in the flux variations at diverse wavelengths, 
since the different-frequency emitting portions of the jet
acquire the same viewing angle at different times 
(Villata \& Raiteri \cite{vil99}).

Radio and optical light-curve behaviours appear quite different, 
the epochs where
the broad radio outbursts are observed not corresponding to the periods when
the faster optical outbursts are seen. However, minor radio flux enhancements
are found simultaneously with the major optical outbursts, as if the mechanism
causing the optical brightenings can only marginally affect the radio
band, while
another process not affecting the optical band is responsible for the
dominant radio events.
Moreover, the characteristic variability time scales of the two bands
are different.

Indeed, cross-correlation
analysis by means of the DCF gives only two weak signals at time lags of
$\sim -887$ days and $\sim 364$ days. 
In the former case radio
variations would lead the optical ones by about 2.4 years, a picture
difficult to explain. The latter possibility, i.e.\
radio variations following the
optical ones by about 1 year, would be more easily interpreted within
the same scenarios which explain the radio-radio delays discussed above.

In any case, it is worth underlying that 
the cross-correlation analysis we performed found only weak correlation
between the optical and radio emissions. This is very different from what 
was found for other BL Lac objects, such
as AO 0235+16 (Raiteri et al.\ \cite{rai01}), which shows strong correlation
between optical and radio bands, with short or even null time lag. 
This is an important difference that models explaining blazar
variability should address.

\begin{acknowledgements}
We would like to thank all astronomers-on-duty and operators at the
IRAM 30 m telescope for their help with the mm-wave observations.
This work was partly supported by the Italian Ministry for University and 
Research (MURST) under grant Cofin 2001/028773 and by the Italian Space Agency 
(ASI) under contract CNR-ASI 1/R/73/01.
This research has made use of:
\begin{itemize}
\item the NASA/IPAC Extragalactic Database (NED), which is operated by the 
Jet Propulsion Laboratory, California Institute of Technology, under
contract with the National Aeronautics and Space Administration;
\item data from the University of Michigan Radio Astronomy Observatory,
which is supported by the National Science Foundation and by funds from
the University of Michigan.
\item observations with the 100 m telescope of the MPIfR
  (Max-Planck-Institut f\"ur Radioastronomie) at Effelsberg.
\end{itemize}
\end{acknowledgements}


\begin{thebibliography}{}

\bibitem[1985]{all85} Aller H.D., Aller M.F., Latimer G.E., Hodge P.E., 1985, 
ApJS 59, 513

\bibitem[1999]{all99}  Aller M.F., Aller H.D., Hughes P.A., Latimer G.A.,
1999, ApJ 512, 601

\bibitem[1986]{ant86} Antonucci R.R.J., Hickson P., Olszewski E.W., Miller
J.S.,  1986, AJ 92, 1

\bibitem[2002]{bac02} Bach U., Krichbaum T.P., Ros E., et al., 2002, in: Ros
E., Porcas R.W., Lobanov, A.P., Zensus J.A. (eds.) Proc.\  6th
European VLBI Network Symposium, p.\ 119 (arXiv:astro-ph/0207083)

\bibitem[2003]{bac03} Bach U., et al., 2003 (in preparation)

\bibitem[1985]{bes85} Beskin G.M., Lyuty{\u\i} V.M., Neizvestny{\u\i} S.I., 
Pustil'nik S.A., Shvartsman V.F., 1985, AZh 62, 432; SvA 29, 252

\bibitem[1979]{bes79} Bessell M.S., 1979, PASP 91, 589

\bibitem[1981]{bie81} Biermann P., Duerbeck H., Eckart A., et al., 1981, ApJ
247,  L53


\bibitem[1994]{blo94} Bloom S.D., Marscher A.P., Gear W.K., et al., 1994, AJ
108,  398


\bibitem[1995]{cle95} Clements S.D., Smith A.G., Aller H.D., Aller M.F., 1995,
AJ 110, 529

\bibitem[1987]{ede87} Edelson R.A., 1987, AJ 94, 1150

\bibitem[1988]{ede88} Edelson R.A., Krolik J.H., 1988, ApJ 333, 646

\bibitem[1998]{gab98} Gabuzda D.C., Kovalev Y.Y., Krichbaum T.P., et al.,
1998, A\&A  333, 445

\bibitem[2000]{gab00} Gabuzda D.C., Kochenov P.Yu., Cawthorne T.V., Kollgaard
R.I., 2000, MNRAS 313, 627

\bibitem[1984]{gea84} Gear W.K., Robson E.I., Ade P.A.R., et al., 1984, ApJ
280,  102

\bibitem[1997]{ghi97} Ghisellini G., Villata M., Raiteri C.M., et al., 1997,
A\&A  327, 61

\bibitem[1999]{gio99} Giommi P., Massaro E., Chiappetti L., et al., 1999, A\&A 
351, 59

\bibitem[2001]{gon01} Gonz\'alez-P\'erez J.N., Kidger M., Mart\'{\i}n-Luis F.,
2001, AJ 122, 2055

\bibitem[1987]{hee87} Heeschen D.S., Krichbaum Th., Schalinski C.J., Witzel
A.,  1987, AJ 94, 1493

\bibitem[1996]{hei96} Heidt J., Wagner S.J., 1996, A\&A 305, 42

\bibitem[1992]{huf92} Hufnagel B.R., Bregman J.N., 1992, ApJ 386, 473

\bibitem[2000]{imp00} Impey C.D., Bychkov V., Tapia S., Gnedin Y., Pustilnik
S., 2000, AJ 119, 1542

\bibitem[2001]{jor01} Jorstad S.G., Marscher A.P., Mattox J.R., et al., 2001,
ApJS 134, 181

\bibitem[1999]{kra99} Kraus A., Quirrenbach A., Lobanov A.P., et al., 1999,
A\&A 344, 807

\bibitem[1981]{kuh81} K\"uhr H., Pauliny-Toth I.I.K., Witzel A., Schmidt J., 
1981, AJ 86, 854

\bibitem[1999]{kur99} Kurtanidze O.M., Nikolashvili M.G., 1999, 
in: Raiteri C.M., Villata M., Takalo
L.O.\ (eds.) Proc.\ OJ-94 Annual Meeting 1999, Blazar Monitoring towards the
Third Millennium. Osservatorio Astronomico di Torino, Pino Torinese, p.\ 25

\bibitem[1996]{mar96} Marchenko S.G., Hagen-Thorn V.A., Yakovleva V.A.,
Mikolaichuk O.V., 1996, PASPC 110, 105

\bibitem[1996]{mas96} Massaro E., Tr\`evese D., 1996, A\&A 312, 810

\bibitem[1999]{mas99} Massaro E., Maesano M., Montagni F., et al., 1999,
PASPC 159, 139 

\bibitem[2002]{nes02} Nesci R., Massaro E., Montagni F., 2002, Publ.\
Astron.\ Soc.\ Aust.\ 19, 143

\bibitem[1994]{ott94} Ott M., Witzel A., Quirrenbach A., et al., 1994,
  A\&A 284, 331

\bibitem[2000]{pen00} Peng B., Kraus A., Krichbaum T.P., Witzel A.,
  2000, A\&AS 145, 1

\bibitem[2001]{pet01} Peterson B.M., 2001, in: The Starburst-AGN
Connection 2001. World Scientific, Singapore (arXiv:astro-ph/0109495)

\bibitem[1998]{pet98} Peterson B.M., Wanders I., Horne K., et al., 1998, PASP
110, 660

\bibitem[1992] {pre92} Press W.H., Teukolsky S.A., Vetterling W.T., Flannery
B.P., 1992, Numerical Recipes in Fortran - The Art of Scientific Computing.
Cambridge University Press, Cambridge

\bibitem[2000]{qia00} Qian B., Tao J., Fan J., 2000, PASJ 52, 1075

\bibitem[2002]{qia02} Qian B., Tao J., Fan J., 2002, AJ 123, 678

\bibitem[1989]{qui89} Quirrenbach A., Witzel A., Krichbaum T., et al., 1989,
Nat 337, 442

\bibitem[1991]{qui91} Quirrenbach A., Witzel A., Wagner S., et al., 1991, ApJ
372, L71 

\bibitem[1992]{qui92} Quirrenbach A., Witzel A., Krichbaum T.P., et al., 1992, 
A\&A 258, 279


\bibitem[1999]{rai99} Raiteri C.M., Villata M., Tosti G., et al., 1999, 
in: Raiteri C.M., Villata M., Takalo
L.O.\ (eds.) Proc.\ OJ-94 Annual Meeting 1999, Blazar Monitoring towards the
Third Millennium. Osservatorio Astronomico di Torino, Pino Torinese, p.\ 76


\bibitem[2001]{rai01} Raiteri C.M., Villata M., Aller H.D., et al., 2001, A\&A
377, 396

\bibitem[2001]{rec01} Rector T.A., Stocke J.T., 2001, AJ 122, 565

\bibitem[1993]{rei93} Reich W., Steppe H., Schlickeiser R., et al., 
1993, A\&A 273, 65

\bibitem[1997]{reu97} Reuter H.-P., Kramer C., Sievers A., et al., 1997, A\&AS 
122, 271

\bibitem[1999]{sag99} Sagar R., Gopal-Krishna, Mohan V., et al., 1999, A\&AS
134,  453

\bibitem[1988]{sil88} Sillanp\"a\"a A., Haarala S., Valtonen M.J., 1988, ApJ
325, 628

\bibitem[1996]{sil96} Sillanp\"a\"a A., Takalo L.O., Pursimo T., et
al., 1996, A\&A 315, L13

\bibitem[1985]{sim85} Simonetti J.H., Codes J.M., Heeschen D.S., 1985, ApJ
296, 46

\bibitem[1995]{smi95} Smith A.G., Nair A.D., 1995, PASP 107, 863

\bibitem[1993]{smi93} Smith A.G., Nair A.D., Leacock R.J., Clements S.D.,
1993, AJ 105, 437

\bibitem[1988]{ste88} Steppe H., Salter C.J., Chini R., et al., 1988, A\&AS
75,  317

\bibitem[1992]{ste92} Steppe H., Liechti S., Mauersberger R., et al., 1992, 
A\&AS 96, 441

\bibitem[1993]{ste93} Steppe H., Paubert G., Sievers A., et al., 1993, A\&AS 
102, 611

\bibitem[1993]{sti93} Stickel M., Fried J.W., K\"uhr H., 1993, A\&AS 98, 393


\bibitem[1994]{tak94} Takalo L.O., Sillanp\"a\"a A., Nilsson K., 1994, A\&AS
107,  497


\bibitem[1998]{ter98} Ter\"asranta H., Tornikoski M., Mujunen A., et al.,
1998,  A\&AS 132, 305

\bibitem[1994]{tor94} Tornikoski M., Valtaoja E., Ter\"asranta H., et al.,
1994, A\&A 289, 673

\bibitem[1992]{val92} Valtaoja E., L\"ahteenm\"aki A., Ter\"asranta H., 1992, 
A\&AS 95, 73

\bibitem[2001]{ven01} Venturi T., Dallacasa D., Orfei A., et al., 2001, A\&A
379, 755 

\bibitem[1999]{vil99} Villata M., Raiteri C.M., 1999, A\&A 347, 30

\bibitem[1998a]{vil98a} Villata M., Raiteri C.M., Lanteri L., Sobrito
G., Cavallone M., 1998a, A\&AS 130, 305

\bibitem[1998b]{vil98b} Villata M., Raiteri C.M., Sillanp\"a\"a A., Takalo
L.O., 1998b, MNRAS 293, L13 


\bibitem[2000]{vil00} Villata M., Mattox J.R., Massaro E., et al., 2000, A\&A
363, 108

\bibitem[2002]{vil02} Villata M., Raiteri C.M., Kurtanidze O.M., et al., 2002,
A\&A 390, 407

\bibitem[1995]{von95} von Montigny C., Bertsch D.L., Chiang J., et al., 1995,
ApJ  440, 525

\bibitem[1996]{wag96} Wagner S.J., Witzel A., Heidt J., et al., 1996, AJ 111,
2187

\bibitem[1991]{wal91} Waltman E.B., Fiedler R.L., Johnston K.J., et al., 1991, 
ApJS 77, 379
 
\bibitem[1992]{wir92} Wiren S., Valtaoja E., Ter\"asranta H., Kotilainen J., 
1992, AJ 104, 1009


\end{thebibliography}
\end{document}